\documentclass[twocolumn]{aastex7}
\usepackage{comment, todonotes}
\usepackage{savesym, amsmath, amstext}
\usepackage[normalem]{ulem}

\usepackage[utf8]{inputenc}
\usepackage{footmisc}

\newcommand{\degree}{\ifmmode {^\circ}\else$^\circ$\fi}

\newcommand{\lx}{\ensuremath{L_{\mbox{\scriptsize X}}}}

\newcommand{\teff}{\ensuremath{T_{\mbox{\scriptsize eff}}}}

\newcommand{\ith}{\ensuremath{^{\rm th}}}
\def\bpminusrp{($G_\mathrm{BP} - G_\mathrm{RP}$)}
\def\gminusk{($G - K$)}

\newcommand{\Msun}{\ifmmode {M_{\odot}}\else{M$_{\odot}$}\fi}
\newcommand{\lapprox }{{\lower0.8ex\hbox{$\buildrel <\over\sim$}}}
\newcommand{\gapprox }{{\lower0.8ex\hbox{$\buildrel >\over\sim$}}}
\newcommand{\halpha}{H$\alpha$}
\def\Ha{$\mathrm{H}\alpha$}

\def\LLX{$L_{\mathrm{X}}/L_{\mathrm{bol}}$}

\def\LLH{$L_{\mathrm{H}\alpha}/L_{\mathrm{bol}}$}
\def\Lbol{$L_{\mathrm{bol}}$}

\def\Prot{$P_{\mathrm{rot}}$}
\def\Ro{$R_\mathrm{o}$}

\def\Lha{$L_{\mathrm{H}\alpha}$}
\def\fx{$f_\mathrm{X}$}

\newcommand{\lessim }{{\lower0.8ex\hbox{$\buildrel <\over\sim$}}}
\newcommand{\Rsun}{\ifmmode {R_{\odot}}\else${R_{\odot}}$\fi}
\newcommand{\Rearth}{\ifmmode {R_{\oplus}}\else${R_{\oplus}}$\fi}
\def\amin{\ifmmode^{\prime}\else$^{\prime}$\fi}
\def\asec{\ifmmode^{\prime\prime}\else$^{\prime\prime}$\fi}

\newcommand{\columbia}{Department of Astronomy, Columbia University, 550 West 120\ith\ Street, New York, NY 10027, USA}
\newcommand{\rutgers}{Department of Physics and Astronomy, Rutgers, the State University of New Jersey, 136 Frelinghuysen Road, Piscataway, NJ 08854, USA}
\newcommand{\westwash}{Department of Physics \& Astronomy, Western Washington University, Bellingham, WA 98225, USA}
\newcommand{\lafayette}{Department of Physics, Lafayette College, Easton, PA 18042, USA}
\newcommand{\montreal}{D\'epartement de Physique, Universit\'e de Montr\'eal, C.P.~6128, Succ.~Centre-Ville, Montr\'eal, Qu\'ebec, CA}
\newcommand{\ou}{Homer L.~Dodge Department of Physics and Astronomy, University of Oklahoma, 440 W.~Brooks St, Norman, OK 73019, USA}
\newcommand{\LAB}{Laboratoire d’astrophysique de Bordeaux, Univ. Bordeaux, CNRS, B18N, Allée Geoffroy Saint-Hilaire, 33615 Pessac, France}

\bibpunct[; ]{(}{)}{,}{a}{}{;}

\shortauthors{Ag{\"u}eros et al.}
\shorttitle{Membership, Age, Rotation, and Activity for Coma Ber}

\begin{document}

\title{Crowning the Queen:\\Membership, Age, Rotation, and Activity for the Open Cluster Coma Berenices}

\correspondingauthor{M.~A.~Ag{\"u}eros}
\email{m.agueros@columbia.edu}

\author[0000-0001-7077-3664]{M.~A.~Ag{\"u}eros}
\affiliation{\columbia}\affiliation{\LAB}
\email{m.agueros@columbia.edu}

\author[0000-0002-2792-134X]{J.~L.~Curtis}
\affiliation{\columbia}
\email{jasoncurtis.astro@gmail.com}

\author[0000-0002-8047-1982]{A.~N\'u\~nez}
\affiliation{\columbia}
\email{alejo.nunez@gmail.com}

\author[0009-0005-0339-015X]{C.~Burhenne}
\affiliation{\rutgers}
\email{cdb201@physics.rutgers.edu}

\author{P.~Rothstein}
\affiliation{\columbia}
\email{ptr1@columbia.edu}

\author{B.~J.~Shaham}
\affiliation{\columbia}
\email{bjs2228@columbia.edu}

\author[0000-0002-7198-5199]{K.~Singh}
\affiliation{\columbia}
\email{ks3693@columbia.edu}

\author[0000-0003-2368-345X]{P.~Bergeron}
\affiliation{\montreal}
\email{pierre.bergeron.1@umontreal.ca}

\author[0000-0001-6098-2235]{M.~Kilic}
\affiliation{\ou}
\email{kilic@ou.edu}

\author[0000-0001-6914-7797]{K.~R.~Covey}
\affiliation{\westwash}
\email{kevin.covey@wwu.edu}

\author[0000-0001-7371-2832]{S.~T.~Douglas}
\affiliation{\lafayette}
\email{douglste@lafayette.edu}

\begin{abstract}
Despite being only 85 pc away, the open cluster \object{Coma Berenices} (Coma Ber) has not been extensively studied. This is due in part to its sparseness and low proper motion, which together made Coma Ber's membership challenging to establish. Gaia data for $\approx$400 previously cataloged candidate cluster stars allowed us to identify $\approx$300 as members. With [Fe/H] measurements for nine members, we found that Coma Ber has a solar metallicity, and then fit isochrones to its color--magnitude diagram to determine that it is 675$\pm$100 Myr old. With photometry obtained by the Transiting Exoplanet Survey Satellite (TESS) and Zwicky Transient Facility (ZTF), we measured rotation periods (\Prot) for 137 of Coma Ber's low-mass stars, increasing the sample of members with measured \Prot\ by a factor of six, and extending the rotational census for the cluster from its late F stars through to its fully convective M dwarfs. By measuring the equivalent width of the \halpha\ line for $\approx$250 stars and collecting X-ray detections for $\approx$100 ($\approx$85\% and $\approx$33\% of the cluster's members, respectively), we characterized magnetic activity in Coma Ber and examined the dependence of chromospheric and of coronal activity on rotation in these stars. Despite having a metallicity that is 0.2 dex below that of their coeval cousins in Praesepe and the Hyades, low-mass stars in Coma Ber seem to follow a similar rotation--activity relation. In detail, however, there are differences that may provide further insight into the impact of metallicity on this still poorly understood relation.
\end{abstract}


\section{Introduction} \label{sect:intro}
The angular-momentum evolution of a low-mass star ($\lapprox$1.2~\Msun) depends on its internal structure, the degree of core-envelope coupling in that interior, the strength of its magnetic field and of the resulting magnetized winds, and the effects exerted by any companions it may have (e.g., through tides or mass transfer). Examining the distribution of rotation periods (\Prot) for such stars in open clusters, whose ages can be accurately determined, provides empirical constraints on how angular momentum changes over time and as a function of mass, and informs theoretical models \citep[e.g.,][]{rebull2004,barnes2007,meibom2009,meibom2015,matt2015,douglas2016,agueros2018,curtis2019,curtis2020,spada2020,godoy2021,bouma2023,vanlane2024}. Furthermore, because of the fundamental connection between rotation and magnetic-field generation, measuring proxies for the field strength of open cluster stars, such as \halpha\ emission, provides  insight into the origin and evolution of magnetic fields in low-mass stars \citep[e.g.,][]{stauffer1997,mohanty2003,mamajek2008,jackson2010,Douglas2014,Nunez2017,Nunez2022,Nunez2024, santos2025}. 

Nature, however, has provided us with a limited number of open clusters whose low-mass members can be examined in this kind of detail. Historically, studies of the age--rotation--activity relation have relied on observations of close-by clusters such as the Hyades, Pleiades, and Praesepe \citep [e.g., the studies of rotation and activity in the Hyades and Pleiades of][]{radick1987, jones1996, stauffer1997}. Spurred by the launch of Kepler  \citep{Borucki2010} and its repurposing as K2 \citep{Howell2014}, and more recently by the use of Gaia photometry and astrometry to establish membership lists far more accurately than was possible previously \citep[e.g.,][]{DR2HRD,CG2018}, investigators have now extensively revisited these benchmark clusters \citep[e.g.,][]{rebull2016a, rebull2016b,douglas2017,douglas2019,Rampalli2021} and extended this work to more distant and challenging open clusters such as NGC 6811 and Ruprecht 147 \citep[e.g.,][]{curtis2019,curtis2020,santos2025}. 

We report here on our work to admit Coma Berenices \citep[also Melotte 111 and Collinder 256, hereafter Coma Ber, R.A.~= 12$^{\rm h}$ 23$^{\rm m}$, decl.~=~25$^\circ$44$\amin$, J2000;][]{CG2020} to the exclusive circle of benchmark open clusters for age--rotation--activity studies. At a distance of only 87~pc \citep[][]{vanleeuwen2009}, Coma Ber is not much farther away than the Hyades, but remains poorly studied, principally because its combination of sparseness and low proper motion made its stars hard to distinguish from the Galactic background \citep{terrien2014,tang2018}. From the time of the first attempt to identify its members in the 1930s \citep{trumpler1938} until the early 2000s, the number of cataloged (candidate) Coma Ber members barely reached 50 stars \citep[e.g.,][]{odenkirchen1998}. And although more recent studies have increased this number, something as fundamental as Coma Ber's age remained uncertain, with literature values ranging from 400~Myr \citep[e.g.,][]{kraus2007b} to 800~Myr \citep[e.g.,][]{tang2018}. 

This paper is structured as follows: in Section~\ref{sec:mem}, we describe how we constructed our Coma Ber membership catalog, which we use in Section~\ref{sec:prop} to obtain the cluster's age. We also discuss in Section~\ref{sec:prop} the age estimates obtained based on the spectra available for the cluster's two known white dwarfs (WDs). In Section~\ref{sec:rot}, we describe our use of optical photometry to measure \Prot\ for Coma Ber stars. In Section~\ref{sec:spec}, we detail our use of optical spectra to measure the strength of the chromospheric \halpha\ line in low-mass members of Coma Ber. In Section~\ref{sec:xray}, we present the coronal X-ray data available for Coma Ber members.

\begin{figure}[!t]
\begin{center}
\includegraphics[trim=0.5cm 0.5cm 0.5cm 0.cm, clip=True, width=1\columnwidth]{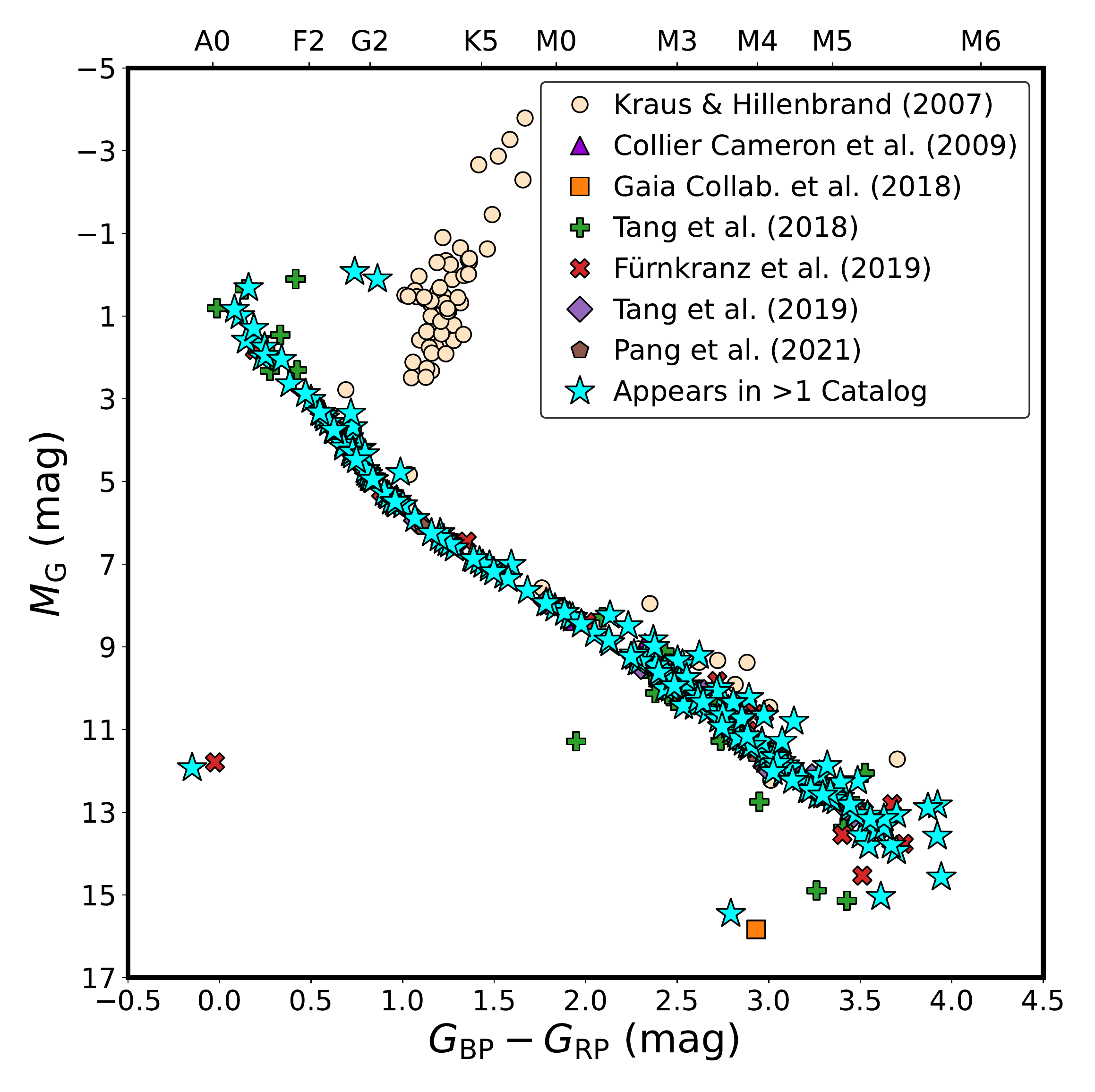}
\vspace{-.6cm}
\caption{CMD for the $\approx$400 candidate members in our initial Coma Ber catalog, a merger of pre-existing catalogs for which we updated the astrometry and photometry to the Gaia DR3 values. Spectral types are indicated along the top axis for reference. The cyan stars appear in more than one of these catalogs. As discussed in the text, this CMD shows important contamination from background giants lacking parallaxes that were included in the \cite{kraus2007b} catalog. These stars are identified as non-members in Figure~\ref{fig:parallax}.} 
\label{fig:orig_cmd}
\end{center}
\end{figure}

\begin{figure}[!t]
\begin{center}
\includegraphics[trim=0.5cm .8cm 0.5cm 0.4cm, clip=True, width=1\columnwidth]{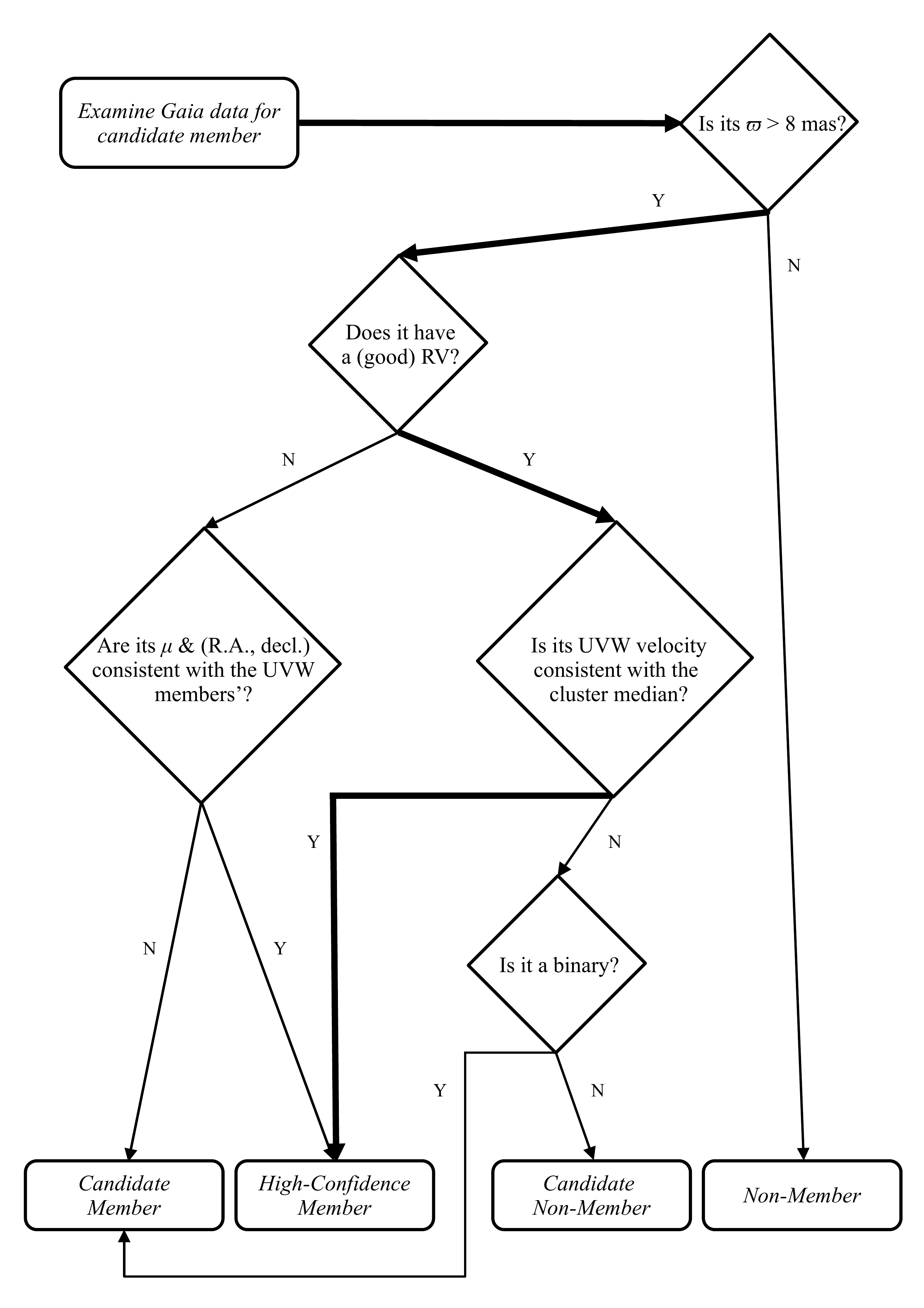}
\caption{Flowchart illustrating our approach to classifying the 379 stars in our initial Coma Ber catalog. We first followed the path indicated by the thick arrows, which allowed us to define a velocity core of 140 high-confidence members for the cluster, before considering the membership status of stars either lacking RVs or whose RVs have large errors ($>$5~km~s$^{-1}$).} 
\label{fig:flow}
\end{center}
\end{figure}

In Section~\ref{sec:disc}, we use these measurements to compare the distributions of rotation periods, \halpha\ equivalent widths, and X-ray luminosities for stars in Coma Ber to those for stars in the benchmark Praesepe and Hyades clusters; we also examine the rotation--activity relation for Coma Ber stars as described by our chromospheric and coronal data. We conclude in Section~\ref{sec:concl}.

\section{Establishing a Gaia-based membership catalog}\label{sec:mem}

\subsection{Merging existing catalogs}
Our original Coma Ber catalog was based on that described in \citet{kraus2007b}. These authors combined data from several large-scale photometric surveys, including SDSS and the Two Micron All Sky Survey \citep[2MASS;][]{2mass}, to identify 149 stars with membership probabilities  $P_{\rm mem} \geq 50\%$, of which 98 have a $P_{\rm mem} > 80\%$. As a basis for our initial catalog, we used an expanded version of the \cite{kraus2007b} catalog that extended to $P_{\rm mem} \geq 30\%$, thereby obtaining as a starting point a sample of 184 stars (A.~Kraus, priv.~comm.). 

We merged this list with the \cite{collier2009} list of candidate members. These authors used data from the SuperWASP optical array \citep{superwasp}, designed to search for transiting exoplanets, to identify more than 1600 rotational variables in the Coma Ber field, for which \cite{collier2009} then measured proper motions and distances from a nominal cluster main sequence in a 2MASS color-magnitude diagram (CMD). The result is a list of 30 F, G, and early K stars with measured \Prot\ and 
$P_{\rm mem} \geq 50\%$, as estimated by these authors.\footnote{A major difficulty in merging these catalogs is how to interpret the differences in $P_{\rm mem}$ for individual stars given the different methods used to calculate these probabilities. Our general approach is to keep even low-probability members in a given catalog until we can complete our own analysis of the full merged catalog.}

We then added to this list the cluster members of \citet{DR2HRD}, \citet{tang2018}, \citet{furnkranz2019}, and \citet{tang2019}, studies that all relied at least in part on the Gaia Data Release~2 \citep{gaiaDR2}, as well as that of \cite{pang2021}, which used the Gaia Early Data Release~3 \citep{edr3}.\footnote{Several additional studies published after our membership analysis was finalized \citep[e.g.,][]{Hunt2024,Liu2025} identify comparable numbers of high-confidence cluster members.} In Table~\ref{tab:lit}, we provide a brief summary of the properties of Coma Ber as described in these different studies, although we caution that some of the values quoted in the table (and in the corresponding papers) were not derived by the authors, or were not always the ones used when e.g., fitting an isochrone to the cluster CMD.\footnote{Specifically, most authors assume a solar metallicity and $E(B-V) = 0$ when fitting an isochrone to the cluster CMD, even when they estimate that the metallicity and/or [Fe/H] is \textbf{not} solar, as the difference from a solar value is generally small.} 

The resulting merged catalog includes 379 stars. The Gaia CMD for these stars, which we considered our initial list of candidate Coma Ber members, is shown in Figure~\ref{fig:orig_cmd}, where we color-code the points by the catalog in which they are originally listed, and provide approximate spectral types for reference.\footnote{\label{note} We used the table  of E.~Mamajek, version 2022.04.16, available at \url{http://www.pas.rochester.edu/~emamajek/EEM_dwarf_UBVIJHK_colors_Teff.txt} to estimate spectral types from Gaia colors. Much of this table comes from \citet{Pecaut2013}.} Of these 379 stars, 211 (56\%) appeared in more than one previous catalog. Below we describe how we proceeded to construct the catalog we used to characterize Coma Ber; a flowchart summarizing the different steps is shown in Figure~\ref{fig:flow}.

\begin{figure}[!t]
\begin{center}
\includegraphics[trim=0.5cm 0.7cm 0.5cm 0.cm, clip=True, width=1\columnwidth]{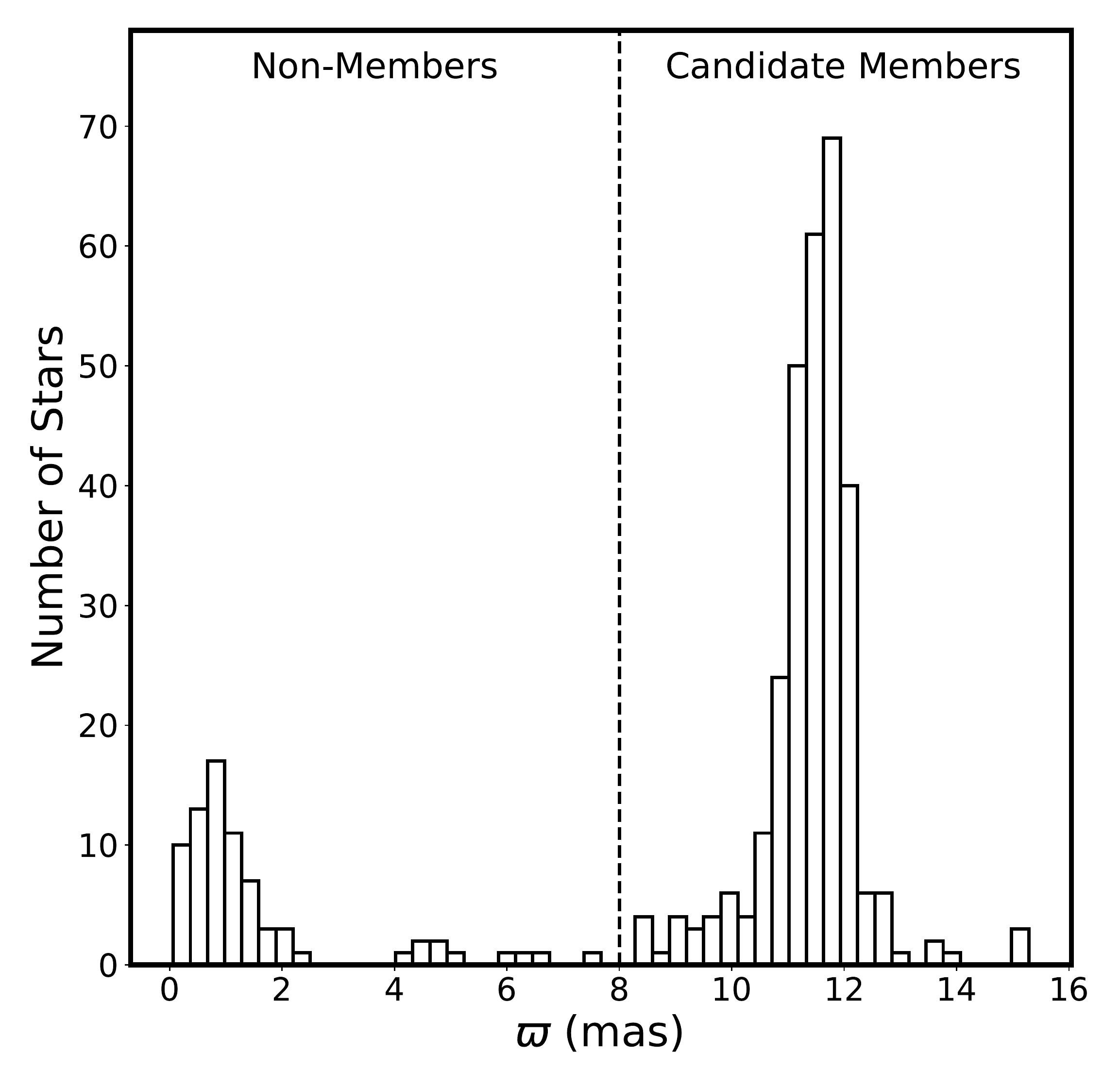}
\caption{An illustration of the discriminating power of Gaia data: DR3 parallax distribution for the 379 stars in our initial Coma Ber catalog. Stars to the left of the $\varpi$~=~8 mas line are likely non-members, with the clump with $\varpi\ \lapprox\ 2$~mas likely to be background giants that contaminated the \cite{kraus2007b} catalog because of the cluster's relatively low proper motion compared to other field stars}.
\label{fig:parallax}
\end{center}
\end{figure}

\begin{deluxetable*}{lDDDcDDcc}[!t]
\centering 

\tablecaption{Catalogs used to identify candidate Coma Ber members \label{tab:lit}}

\tablehead{
\colhead{Authors} &
\multicolumn2c{$\mu_\alpha$ cos $\delta$} & 
\multicolumn2c{$\mu_\delta$} &
\multicolumn2c{D} &
\colhead{Age} &
\multicolumn2c{$E(B-V)$} &
\multicolumn2c{[Fe/H]} &
\colhead{Total} & 
\colhead{Unique} \\[-0.1in]
  &
\multicolumn2c{(mas yr$^{-1}$)} &
\multicolumn2c{(mas yr$^{-1}$)} &
\multicolumn2c{(pc)} &
\colhead{(Myr)} &
\multicolumn2c{(mag)} &
\multicolumn2c{(dex)} & \colhead{Members}
 & \colhead{Members\tablenotemark{a}}
}
\decimals
\startdata
\cite{kraus2007b} & $-$10.30\tablenotemark{b} & $-$8.80\tablenotemark{b} & 90.0 & 400 & 0.00\tablenotemark{c} & \multicolumn2r{\nodata} & 184\tablenotemark{d} & 82 \\
\citet{collier2009} & $-$11.70 & $-$7.95 & 89.9\tablenotemark{e} & 520 & 0.006\tablenotemark{f} & 0.00 & 30\tablenotemark{g} & 4 \\
\cite{DR2HRD} & $-$12.11 & $-$9.00 & 85.9 & 646 & 0.00 & 0.00\tablenotemark{h} & 153 & 3 \\
\cite{tang2018} & $-$12.00 & $-$9.05 & 86.7\tablenotemark{i} & 800 & 0.006\tablenotemark{f} & 0.00\tablenotemark{j} & 192\tablenotemark{k} & 33 \\
\cite{furnkranz2019} & $-$11.84 & $-$8.69 & 85.0 & 700 & \multicolumn2c{\nodata} & $-$0.12 & 214 & 20\\
\cite{tang2019} & $-$11.80 & $-$8.66 & 85.8\tablenotemark{l} & 700--800 & 0.00\tablenotemark{m} & $-$0.02 & 197 & 23 \\
\cite{pang2021} & \multicolumn4c{\nodata} & 86.4 & 700\tablenotemark{n} & 0.00\tablenotemark{n} & \multicolumn2r{\nodata \hspace{-.2cm}\tablenotemark{o}}  & 158 & 3\\
\enddata
\tablecomments{Most authors took as a starting point for their search for members $\mu_{\alpha}$ cos $\delta$ = $-$11.21 mas yr$^{-1}$ and $\mu_{\delta}$ = $-$9.16 mas yr$^{-1}$, as found by \citet{vanleeuwen1999} using Hipparcos data. The quoted proper motions are the median values for members in the individual catalogs. \cite{DR2HRD} do not provide proper motions for individual cluster members in their catalog, and instead give solutions from their fits for the motions and distance. Finally, the quoted ages are generally from isochrone fits assuming a solar metallicity even if the authors found that Coma Ber has a non-solar metallicity.}
\tablenotetext{a}{Stars  included in our initial merged catalog of candidate members that \textbf{do not} appear in any of the other individual catalogs.} 
\tablenotetext{b}{For stars with $P_{\rm mem} \geq 50\%$.}
\tablenotetext{c}{Adopted from \cite{feltz1972}.}
\tablenotetext{d}{\cite{kraus2007b} list 149 members with a $P_{\rm mem} \geq 50\%$; we expanded the sample to $P_{\rm mem} \geq 30\%$.}
\tablenotetext{e}{Adopted from \cite{vanleeuwen1999}.}
\tablenotetext{f}{Adopted from \cite{nicolet1981}.}
\tablenotetext{g}{With $P_{\rm mem} > 50\%$.}
\tablenotetext{h}{Adopted from \citet{netopil2016}. \cite{DR2HRD} estimate a metallicity $Z = 0.017$ dex based on their isochrone fit to the CMD.}
\tablenotetext{i}{For stars with Gaia DR2 parallaxes.}
\tablenotetext{j}{Adopted from \cite{friel1992}.}
\tablenotetext{k}{\cite{tang2018} estimate that 23\% of these stars are not true members.}
\tablenotetext{l}{For the 77 stars within the $\approx$7 pc tidal radius.}
\tablenotetext{m}{Adopted from \citet{casewell2006}.}
\tablenotetext{n}{Adopted from \cite{tang2019}.}
\tablenotetext{o}{\cite{pang2021} quote $Z = 0.015$ dex, which appears to originate from the \cite{furnkranz2019} isochrone fit.}

\vspace{-0.25in}

\end{deluxetable*}

\begin{figure}[t]
\begin{center}
\includegraphics[trim=1.5cm 1.6cm .8cm 1.cm, clip=True, width=1.0\columnwidth]{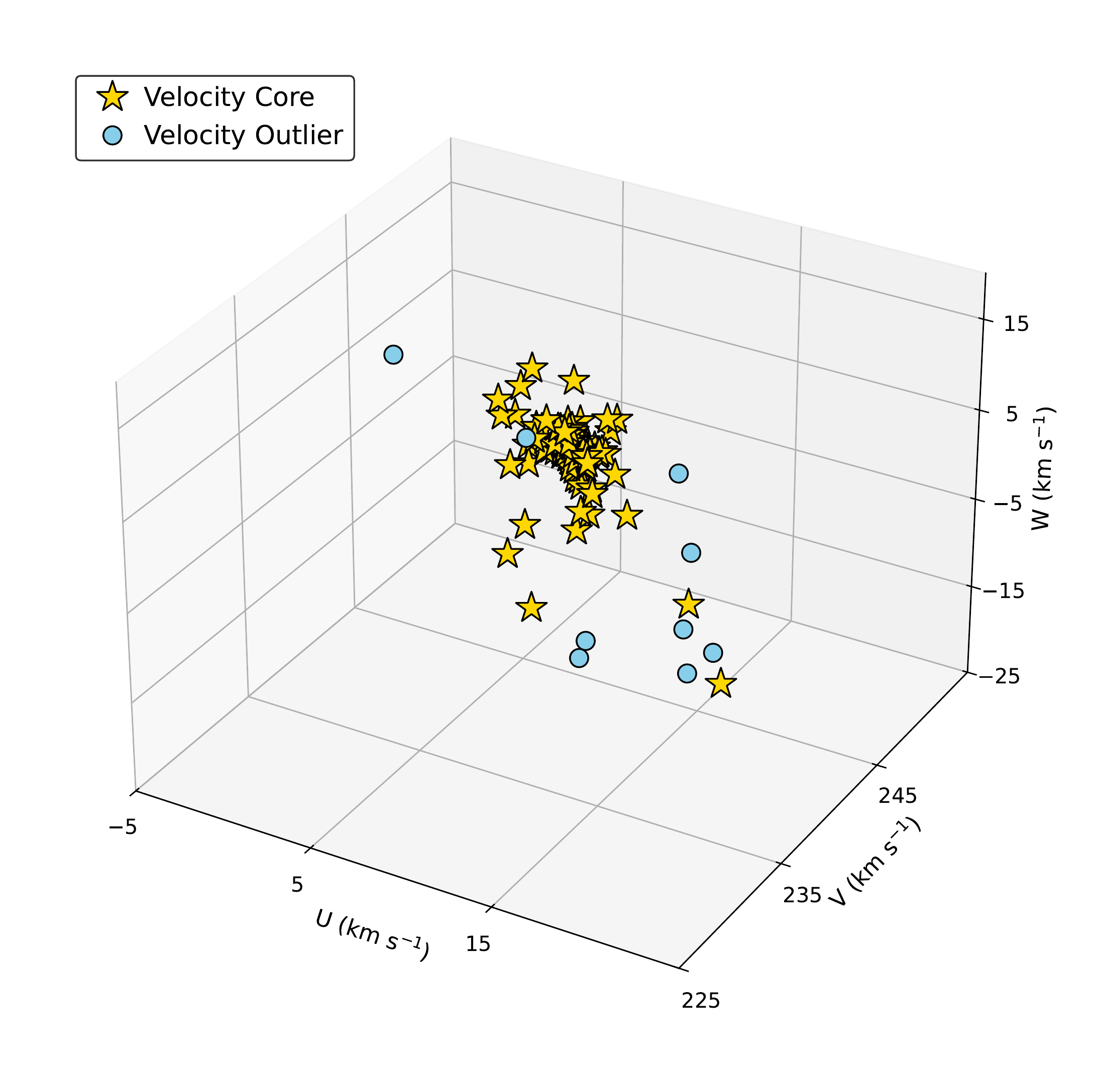}
\caption{U, V, and W velocities for the 181 stars with $\varpi$~$>$~8 mas and Gaia RV measurements in our initial catalog. The gold stars are the 140 stars that form the cluster's velocity core and are classified as high-confidence members, while the nine light blue points are UVW outliers and therefore candidate non-members. We used the proper motions of the 140 high-confidence members to select members from the 119 stars with $\varpi$~$>$~8 mas but lacking RVs, and from the 24 stars having RVs with RV$_{\rm err} > 5$~km~s$^{-1}$ (see Figure~\ref{fig:pms}).}
\label{fig:uvw}
\end{center}
\end{figure}

\subsection{Using Gaia data to identify high-confidence and candidate members}

Our next step was to vet these candidate members using the Gaia Data Release 3 \citep[DR3;][]{GaiaDR3}. The power of Gaia data to clean a membership catalog is illustrated in Figure~\ref{fig:parallax}, where we show the Gaia DR3 parallax ($\varpi$) distribution for our candidate members. We found that $\approx$20\% of our candidates have a $\varpi$ that places them well beyond the $\lapprox$100~pc distance to the cluster. This includes a significant number with 0~$<$~$\varpi$~$<$~2~mas that are likely to be background giants. These stars contaminated the \citet{kraus2007b} catalog, constructed without parallaxes, because, despite its proximity, Coma Ber has a comparatively low proper motion relative to other field stars. We therefore started by requiring that candidate cluster members have $\varpi$~$>$~8~mas, thereby eliminating 79 background giants and other distant, contaminating stars. 

We analyzed the remaining 300 stars using Gaia DR3 radial velocities (RVs), proper motions, and positions. First, we calculated Cartesian U, V, and W velocities for the 181 stars with RVs. We flagged the 24 stars with RV uncertainties RV$_{\rm err}>5$~km~s$^{-1}$; the membership status of these stars was considered along with that of stars lacking RV measurements, and is discussed below. To identify a core of co-moving cluster members, we then calculated $\Delta$UVW, the difference between the total UVW velocity of a candidate member (i.e., $\sqrt{{\rm U}^2+{\rm V}^2+{\rm W}^2}$) and the median total UVW velocity for the 157 stars with small RV$_{\rm err}$. 

That $\Delta$UVW for a {\it bona fide} member of an open cluster be no more than a few km~s$^{-1}$ is an expectation from theory. For example, \citet[][]{bate2012} found that the velocity dispersion for stars in a simulated cluster $\approx$10$^5$~yr after their natal cloud began to collapse is of order 5~km~s$^{-1}$, and varied by $\pm$1--2~km~s$^{-1}$ depending on mass and binary status. Meanwhile, Gaia data for loosely bound open clusters indicate that the typical RV dispersion is of order 1~km~s$^{-1}$ \citep{soubiran2018}, although searches in Gaia data for larger-scale co-moving populations have identified candidate structures with larger dispersions \citep[e.g.,][]{kounkel2019}.\footnote{Whether \textbf{all} of these structures do in fact contain physically associated stars is debated \citep[cf.~discussion in][]{zucker2022}, but follow-up studies of several have shown convincingly that their stars are co-eval, co-chemical, and co-moving, strongly suggesting a common origin \cite[e.g.,][]{bouma2021, andrews2022, tregoning2024, miller2025}. See \cite{cg2022} for a review of Gaia's transformation of our view of star clusters in the Milky Way.}

\begin{figure}[!th]
\begin{center}
\includegraphics[trim=0.5cm 0.5cm 0.5cm 0.1cm, clip=True, width=1\columnwidth]{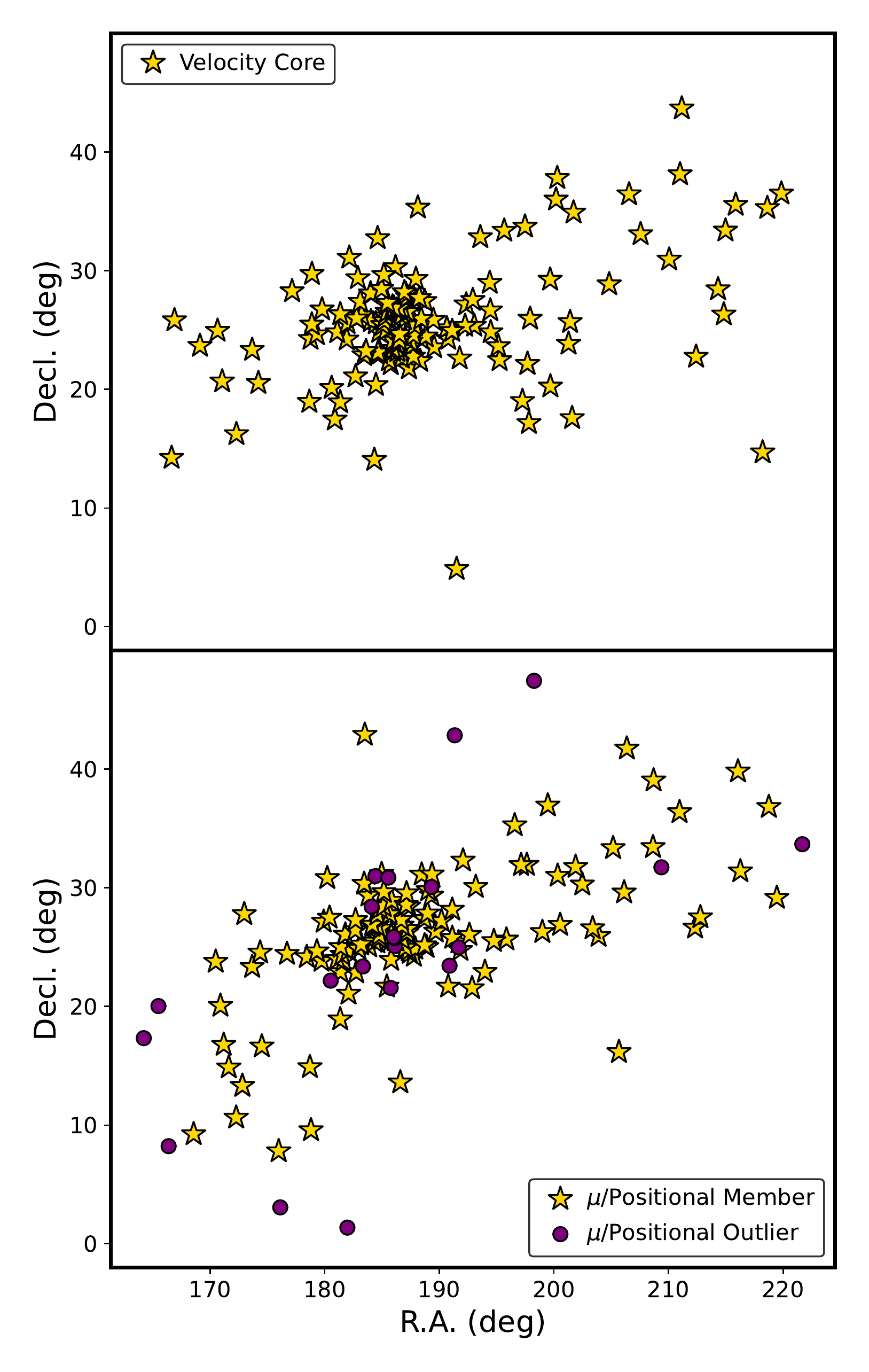}
\caption{\textit{Top---}Positions for the 140 stars confirmed as high-confidence members of Coma Ber based on their UVW velocities (see Figure~\ref{fig:uvw}). We used this distribution to test the membership of the 143 stars with $\varpi$~$>$~8 mas but lacking or having unreliable RVs in our sample. \textit{Bottom---}Positions for those 143 stars. The 122 stars whose positions and proper motions all fell within the range defined by the astrometry for the UVW high-confidence members were also classified as high-confidence members and are shown as gold stars. The 21 purple points are positional and/or proper-motion outliers, and are classified as candidate members until their RVs can be obtained.} 
\label{fig:pms}
\end{center}
\end{figure}

From this set of stars, we therefore selected those for which $|{\rm \Delta UVW}|$~$\leq$~2~km~s$^{-1}$ or for which $\Delta$UVW/RV$_{\rm err} \leq$~1. We found that 140 of the 157 stars (89\%) meet these criteria, and we labeled these stars velocity core members (see Figure~\ref{fig:uvw}). The median velocities of these 140 stars, which we classified as high-confidence cluster members, are 
U = 10.52$\pm$0.11, 
V = 240.02$\pm$0.06, and
W = 7.20$\pm$0.33~km~s$^{-1}$.
Of these high-confidence members, 38 (27\%) are known or possible binary stars (see Section~\ref{sec:bin}).

While the 17 stars that do not meet these $\Delta$UVW criteria are velocity outliers, some may still be members of Coma Ber. In particular, binary members may have large RV offsets relative to the cluster median and show no RV variability. We therefore searched for known and candidate binaries among these 17 velocity outliers. 

We found two known single-lined spectroscopic binaries, \object{Gaia DR3 3960724286666918016} and \object[Gaia DR3 3960724252307532672]{3960724252307532672}, and we used the Gaia re-normalized unit weight error (RUWE) to identify six stars for which RUWE $>$ 1.4 as possible binaries (see discussion in Section~\ref{sec:bin}). We classified these eight known/possible binaries as candidate Coma Ber members, as additional RV data are required to determine whether they belong to the cluster. And we classified the other nine UVW outliers as candidate non-members.

\begin{figure}[!t]
\begin{center}
\includegraphics[trim=0.5cm 0.5cm 0.5cm 0.cm, clip=True, width=1\columnwidth]{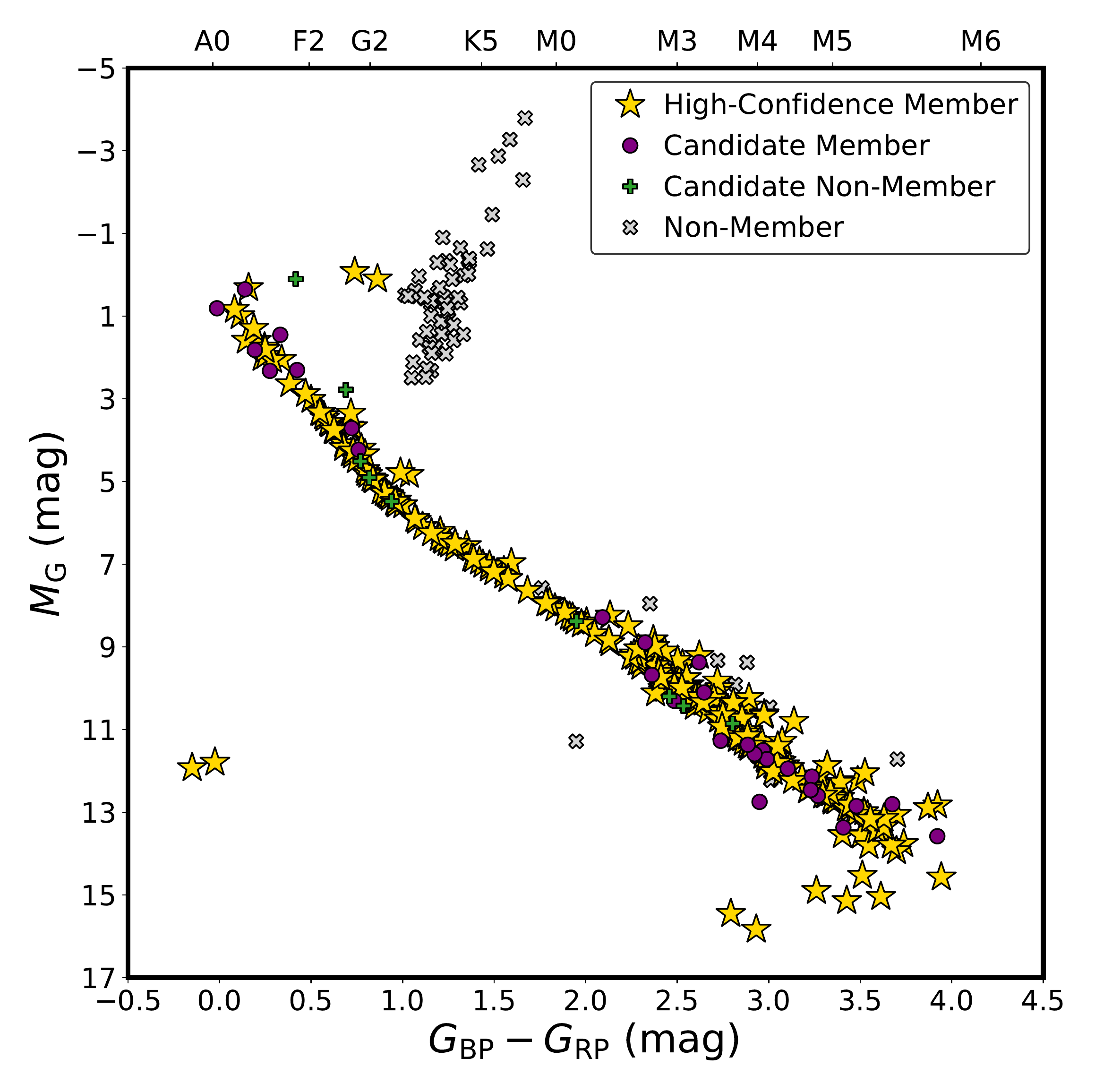}
\caption{Updated Gaia CMD for Coma Ber. Using velocity and astrometric cuts, we identify 262 stars as high-confidence members (gold stars), 29 stars as candidate members (purple points), nine as likely non-members (green pluses), and 79 as non-members (gray crosses).} 
\label{fig:clean_cmd}
\end{center}
\end{figure}

Having completed our velocity-based analysis, we assumed that the 140 velocity core members define the extent of Coma Ber in both $\mu$ and (R.A., decl.) space. We then assigned the 143 stars with $\varpi$~$>$~8~mas that lack RVs (119 stars) or have RVs with relatively large uncertainties (the 24 stars mentioned above) to a membership category based on how their proper motions and positions on the sky compared to those of the velocity core members. The velocity core's stars have $-18.62 \leq \mu_\alpha {\rm cos}\ \delta \leq -1.52$ and $-23.91 \leq \mu_\delta \leq -4.9$ mas yr$^{-1}$, and $166.64 \leq$~R.A.~$\leq 219.85$ and $4.87 \leq$~decl.~$\leq 43.68^\circ$. To be considered a high-confidence member, stars lacking an RV or with a large RV uncertainty needed to have proper motions \emph{and} positions falling within those same ranges. 

Of these 143 stars, 122 (85\%) were classified as high-confidence members based on their two-dimensional (2D) astrometry (see Figure~\ref{fig:pms}). This high percentage of members confirmed that a high-quality set of Coma Ber members can be identified using only parallaxes, proper motions, and positions on the sky. We considered the 21 stars that lack RVs and are proper motion and/or positional outliers to be candidate members, as again, RV data are required to determine their membership status.

The result of this analysis is a sample of 262 high-confidence Coma Ber members identified based on their parallaxes and either their UVW velocities or their proper motions and positions on the sky. The median proper motions for these high-confidence members are $\mu_\alpha$~cos~$\delta = -11.86$~mas~yr$^{-1}$ and $\mu_\delta = -8.74$~mas~yr$^{-1}$, consistent with the median values obtained from the studies listed in Table~\ref{tab:lit} of ($-$11.80, $-$8.69)~mas~yr$^{-1}$. 

Our final catalog also features 29 candidate members, for which RV data are required, and nine candidate non-members. Details about these 300 stars, all of which passed our initial $\varpi$ cut, are included in our membership catalog; Table~\ref{tab:cat} lists the columns in this catalog, available online. The updated Coma Ber CMD is shown in Figure~\ref{fig:clean_cmd}. 

In the analysis that follows, unless otherwise specified, we use ``cluster members'' to refer to the total of 291 high-confidence and candidate members included in our membership catalog.

\begin{figure}[!t]
\begin{center}
\includegraphics[trim=0.3cm 0.3cm 0.3cm 0.cm, clip=True, width=1\columnwidth]{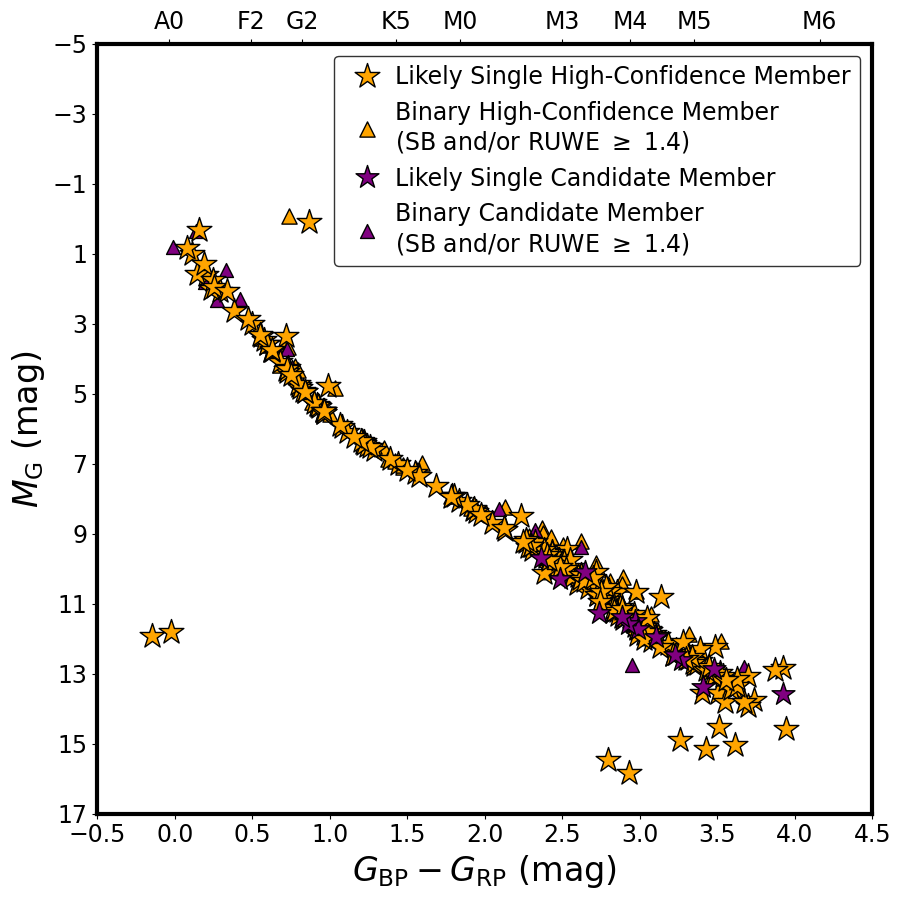}
\caption{Coma Ber CMD showing likely single stars (star symbols) and known/possible binaries (triangles). The orange symbols are high-confidence members, and purple ones are candidate members. In addition to spectroscopic binaries (SBs) identified in the literature, we flag as possible binaries stars for which the Gaia RUWE $\geq$ 1.4. In total, 230 of our Coma Ber members are likely single stars, and 61 are spectroscopic and/or candidate binaries.} 
\label{fig:binaries}
\end{center}
\end{figure}

\subsection{Flagging known and possible binary members}\label{sec:bin}

The presence of a companion can impact the evolution of a star, thereby complicating the use of isochrones to extract a population age, or the interpretation of rotational and/or activity measurements. As a result, we flagged the known binaries in our catalog and searched for stars with Gaia RUWE values $>$1.4. Such high RUWE values are an indication that the astrometric parameters of a source are degraded, and generally imply that it is an unresolved binary \citep[e.g.,][]{Belokurov2020, stassun2021}.

We identified 12 of the high-confidence members as known binaries, and an additional 35 as possible binaries based on their RUWE (only one of the 12 known binaries has a RUWE~$<$~1.4). Of the 29 candidate members, two are known binaries, and 12 more have RUWE~$>$~1.4. These stars are flagged in our final membership catalog.

Figure~\ref{fig:binaries} is a CMD in which we distinguish between likely single stars on one hand and spectroscopic and candidate binaries on the other, and also color-code the stars by whether they are high-confidence or candidate cluster members. Of the 291 Coma Ber members we identified, 230 are classified as likely single stars (79\%) and 61 as spectroscopic and/or candidate binaries (21\%). If we consider just the 262 high-confidence members, the binary fraction is 18\%, and it is 48\% for the candidate members. 

\subsection{Calculating masses and bolometric luminosities}\label{sec:masses}
We derived masses $M_\star$ and bolometric luminosities \Lbol\ from 2MASS $K$ magnitudes for all the main-sequence stars in our catalog. We first converted apparent $K$ magnitudes to absolute $M_K$ magnitudes using our Gaia DR3 parallax-derived distance for each star.\footnote{$M_K$ has long been used to estimate masses for low-mass stars because of its weak sensitivity to metallicity; see, e.g., \cite{baraffe98, delfosse2000, mann2015-1}.} Next, we used empirical $M_K-M_\star$ and $M_K -$ log(\Lbol) relations\footnote{\label{note2} Derived from the table of E.~Mamajek, version 2022.04.016. See Footnote~\ref{note}.} to obtain $M_\star$ and log(\Lbol). These quantities are also included in our final membership catalog. 

\begin{deluxetable}{@{}ll}
\tabletypesize{\footnotesize} 

\tablecaption{Columns in our Coma Ber membership catalog \label{tab:cat}}

\tablehead{
\colhead{Column} & \colhead{Description} 
}

\startdata
1      & Gaia DR3 designation \\
2      & 2MASS designation \\
3, 4   & R.A., Decl. for epoch J2000\\
5      & Membership status\tablenotemark{a} \\
6, 7   & $\varpi$ and 1$\sigma$ uncertainty \\
8, 9   & $\mu_\alpha$ cos $\delta$ and 1$\sigma$ uncertainty \\
10, 11 & $\mu_\delta$ and 1$\sigma$ uncertainty \\
12, 13 & Distance and 1$\sigma$ uncertainty \\
14, 15 & Radial velocity and 1$\sigma$ uncertainty \\
16--18 & Cartesian U, V, and W velocities \\
19, 20 & Gaia DR3 $G$  magnitude and 1$\sigma$ uncertainty \\
21     & Gaia DR3 \bpminusrp \\
22, 23 & 2MASS $K$  magnitude and 1$\sigma$ uncertainty \\
24     & Binary flag: (0) no binary flag; (1) candidate binary; (2) confirmed binary \\
25     & Gaia DR3 RUWE \\
26     & Rotation period \Prot \\
27     & Source of \Prot\tablenotemark{b} \\
28     & Stellar mass $M_*$ \\
29     & Convective turnover time $\tau$ \\
30     & Rossby number \Ro \\
31, 32 & Measured \halpha\ equivalent width EW and 1$\sigma$ uncertainty \\
33     & Relative \halpha\ EW \\
34, 35 & Effective temperature \teff\ and 1$\sigma$ uncertainty \\
36, 37 & $\chi$ and 1$\sigma$ uncertainty \\
38, 39 & \LLH\ and 1$\sigma$ uncertainty \\
40, 41 & X-ray energy flux \fx\ (0.1--2.4 keV) and 1$\sigma$ uncertainty \\
42, 43 & \Lbol\ and 1$\sigma$ uncertainty \\
44, 45 & \LLX\ and 1$\sigma$ uncertainty \\
\enddata
\tablenotetext{a}{The 300 stars in our final membership catalog include 262 high-confidence members (HCM), 29 candidate members (CM) and nine candidate non-members (CNM).}
\tablenotetext{b}{C09: \citet{collier2009}; T14: \citet{terrien2014}; T: TESS; Z: ZTF.} 

\vspace{-0.25in}

\end{deluxetable}

\section{Cluster properties}\label{sec:prop}
\subsection{Obtaining a  metallicity and a reddening}
J.~Brewer observed solar-type members of Praesepe and of Coma Ber with Keck HIRES \citep{HIRES} and provided us with a table of stellar properties derived with Spectroscopy Made Easy \citep[SME;][]{sme} using the same line list and following the same procedure as in \citet{Brewer2016}.\footnote{J.~Brewer, priv.~comm.; see also \citet{Brewer2015}.} Restricting this sample to likely single stars,\footnote{Our membership catalog for Praesepe, which includes information on multiplicity, is described in \citet{Nunez2022}.} we find that based on 10 members, Praesepe has [Fe/H] = +0.189$\pm$0.017~dex, and that based on nine members, Coma Ber has [Fe/H] = +0.003$\pm$0.015~dex.

This [Fe/H] for Praesepe is consistent with a number of recent determinations using different approaches. For example, \cite{gebran2019} applied principal component analysis and the BACCHUS spectral analysis code \citep{masseron2016} to derive an [Fe/H] for three G members of the cluster of 0.17$\pm$0.07 dex. \cite{dorazi2020} found an average [Fe/H]~=~0.21$\pm$0.01 while using a line-by-line approach to derive stellar properties and measure elemental abundances for a sample of 10 solar-type Praesepe members---an approach similar to that of \cite{Vejar2021} for five F and G members with Keck HIRES spectra, which resulted in an [Fe/H]~=~0.21$\pm$0.02 dex. And the result for Coma Ber is in agreement with previous findings that Coma Ber's [Fe/H] is solar to slightly sub-solar (see Table~\ref{tab:lit}).

We used the HIRES sample of Coma Ber stars observed by J.~Brewer, and a large selection of hundreds of nearby solar-type stars observed and analyzed in the same fashion \citep{Brewer2016}, to determine the interstellar reddening in the direction of the cluster. We followed the approach described in appendix A of \citet[][]{douglas2019}: we compared the spectroscopically derived effective temperatures (\teff), which are not sensitive to reddening, to the photometrically derived \teff\ obtained from Gaia colors, which are.

After filtering out giants and likely binaries, we fit a temperature--metallicity--color relation to the \citet{Brewer2016} data that allows us to predict intrinsic Gaia colors from spectroscopic \teff\ and [Fe/H]. We then computed the color excess by subtracting the predicted intrinsic colors from the observed Gaia colors, and converted these values to visual extinctions using Gaia's extinction coefficients.\footnote{\url{https://www.cosmos.esa.int/web/gaia/edr3-extinction-law}} We find $A_V = 0.004$~mag, an unsurprising result given Coma Ber's proximity to Earth and location within the Local Bubble.

\begin{figure}[!t]
\begin{center}
\includegraphics[trim=0.5cm 0.5cm 0.5cm 0.cm, clip=True, width=1\columnwidth]{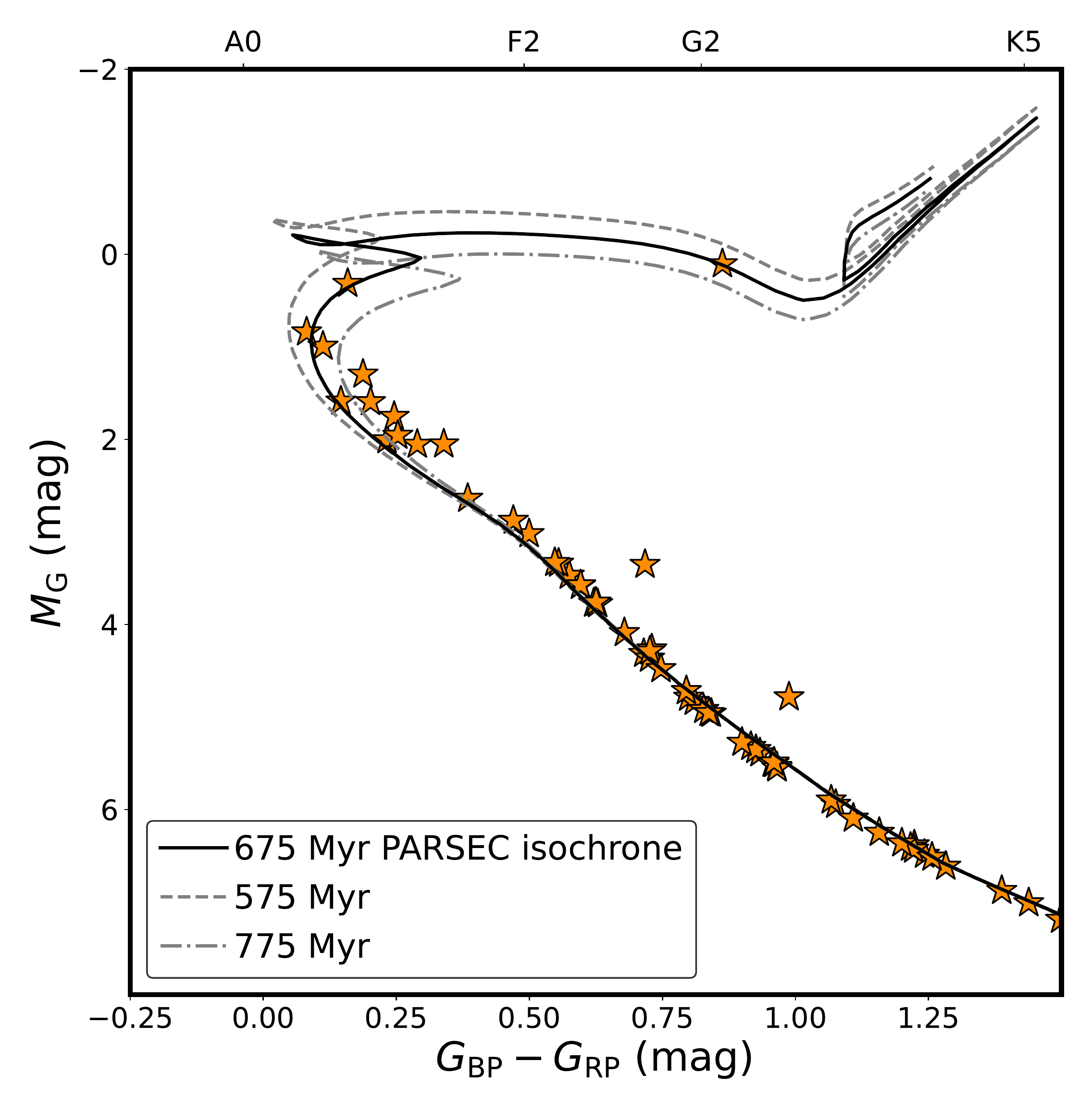}
\caption{PARSEC 675-Myr isochrone overlaid on the CMD for likely single high-confidence members of Coma Ber. We zoom in on the main-sequence turnoff, where the distinction between the different isochrone predictions are most evident.}
\label{fig:isochrone}
\end{center}
\end{figure}

\subsection{Estimating an isochrone age for Coma Ber}\label{sec:iso}

To determine an isochrone age for Coma Ber, we used the distance moduli for high-confidence, likely single members obtained using their inverse Gaia parallaxes. Given that $A_V$ = 0.004 mag for the cluster, we did not apply an extinction correction.

We then fit PAdova and TRieste Stellar Evolution Code (PARSEC) isochrones \citep{bressan2012} to our new CMD for the cluster. To optimize our fit, we varied the metallicity of the PARSEC isochrones while fitting the solar analogs in the cluster. In this manner, we determined that the cluster's  [M/H]~=~$-0.04$ dex; we also derived a value of +0.16~dex for Praesepe. The resulting difference  of $-0.20$ dex from this isochrone-fitting exercise is consistent with the $-0.19$~dex difference in [Fe/H] from the HIRES results. 

With age the only remaining free parameter, we visually fit the turnoff and subgiant members of the cluster. The best fit is for a 675~Myr isochrone, as shown in Figure~\ref{fig:isochrone}; we estimated the uncertainty of this age to be $\pm$100~Myr. This age is consistent with the ages found by the more recent (post-Gaia DR2) studies in Table~\ref{tab:lit}.

We previously estimated an age of 670$\pm$67 Myr for Praesepe \citep[cf.~discussion in][]{douglas2019}. As a sanity check, we refitted the Praesepe DR3 CMD in a similar fashion as we did for Coma Ber, and found an age consistent with this earlier estimate. We concluded that the clusters are coeval.  

\subsection{Estimating an age based on (one of) Coma Ber's WDs}

The CMD in Figure~\ref{fig:clean_cmd} features two WDs with \bpminusrp\ $\approx0$ and $M_{\rm G}\approx12$ mag, both of which are high-confidence members of the cluster. Crucially, each has an archival spectrum available, one obtained by the Sloan Digital Sky Survey \cite[SDSS;][]{york00}, and the other with the Large Sky Area Multi-Object Fiber Spectroscopic Telescope \citep[LAMOST;][]{lamost1}.
These WDs provide another opportunity for estimating the cluster's age. Models fit to their spectra can determine their masses and cooling ages ($\tau_{\rm cool}$). These masses and an initial-to-final mass relation (IFMR) can be used to find the stars' main-sequence (initial) masses, which in turn provide an estimate of the stars' main-sequence lifetimes. Finally, the main-sequences lifetimes can be added to the $\tau_{\rm cool}$ values to find the stars', and by extension, Coma Ber's, age. 

One of the two WDs, \object{Gaia DR3 4008511467191955840} (SDSS J121856.17$+$254557.0), is included in the \citet{kilic2020} study of WDs within 100 pc. These authors model the WD's fundamental parameters based on its SDSS and Panoramic Survey Telescope and Rapid Response System \citep[Pan-STARRS;][]{panstarrs} photometry and its Gaia DR2 parallax. To test the model's reliability, the corresponding  spectrum is then compared to the WD's SDSS spectrum. In this manner, \citet{kilic2020} determined that this is a DA3.1 WD, with a $\teff =  16,099$$\pm$225~K and a mass of 0.90$\pm$0.01~\Msun, and that its $\tau_{\rm cool}$~=~342$\pm$15~Myr. 

Our analysis of the data available for the second WD, \object{Gaia DR3 4001560041148002432} (LAMOST J121100.25$+$225221.2), was less satisfying. We include in Appendix~\ref{app:wds} the output from this analysis, which followed that outlined in \citet{kilic2020}. This, however, resulted in a best-fitting pure-hydrogen atmosphere model that predicted a much broader H$\alpha$ line than seen in the WD's actual spectrum (see Figure~\ref{fig:app_wd1}). 

We therefore fit the normalized Balmer lines to obtain another estimate of the WD's parameters (see Figure~\ref{fig:app_wd2}). The resulting values differed significantly from those obtained from the photometry. The WD's mass is 0.638~\Msun\ and its $\tau_{\rm cool}$~=~408~Myr photometrically, but 0.838~\Msun\ and 814~Myr spectroscopically. 

\begin{figure*}[!th]
\centerline{\includegraphics[trim=0.33cm 0.2cm 0.33cm 0.3cm, clip=True, width=2.1\columnwidth]{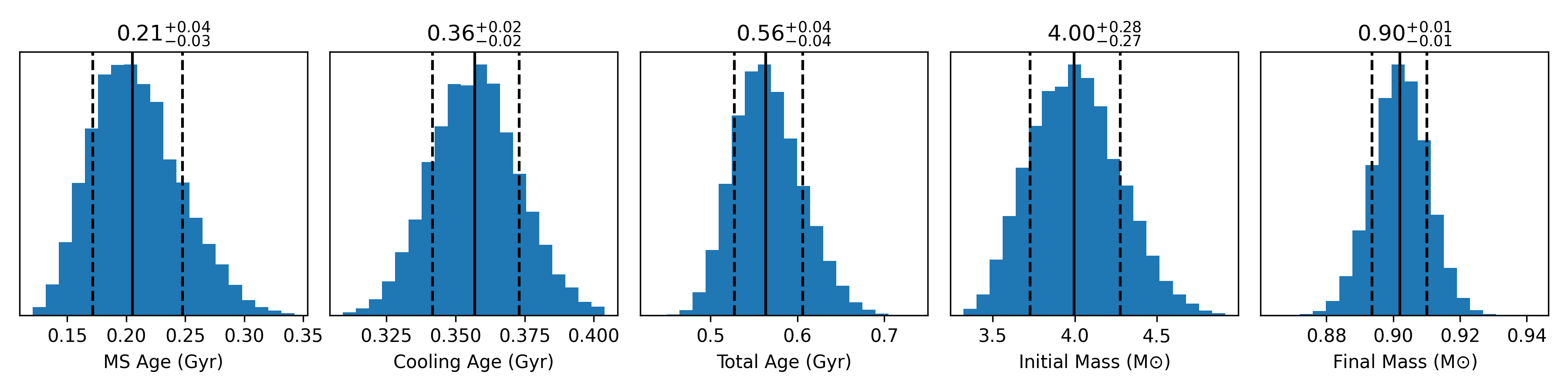}}
\caption{{\tt wdwarfdate} outputs for \object{Gaia DR3 4008511467191955840}: its progenitor's main-sequence age, its cooling age, its total age, its progenitor's initial mass, and its mass.}
\label{fig:wdwarfdate}
\end{figure*}

Resolving the discrepancies between the photometric and spectroscopic solutions for this WD is beyond the scope of this paper. We note, however, that such discrepancies can signal the presence of a double-degenerate system \citep[e.g.,][]{bedard2017,kilic2020b}. In such a system, a cooler companion can add extra continuum flux and make the WD appear brighter, resulting in a lower mass estimate in the photometric fit. Regardless, we set aside this WD and focused on obtaining an age for \object{Gaia DR3 4008511467191955840}.

To do so, we turned to {\tt wdwarfdate} \citep{wdwarfdate}, which derives the Bayesian age of a WD from its \teff\ and log g.
Crucially, this code incorporates all of the different steps mentioned in the first paragraph of this section. We used {\tt wdwarfdate} with the \citet{bedard2020} WD cooling model, the \citet{cummings2018} IFMR, and the Modules for Experiments in Stellar Astrophysics \citep[MESA;][]{Choi2016} single-star evolution model to estimate the mass and $\tau_{\rm cool}$ of \object{Gaia DR3 4008511467191955840}, the mass and lifetime of its progenitor star, and the total age of the object, along with the relevant uncertainties.

{\tt wdwarfdate}'s outputs are  shown in Figure~\ref{fig:wdwarfdate}. As indicated by the middle panel, the code favors an age of 560$\pm$40 Myr. This implies that Coma Ber is younger than what we found from our isochrone fit, but the ages are consistent within the uncertainties. Ultimately, we considered this further evidence that Coma Ber is about the same age as Praesepe.

\begin{figure*}[!t]
\centerline{\includegraphics[width=1.99\columnwidth]{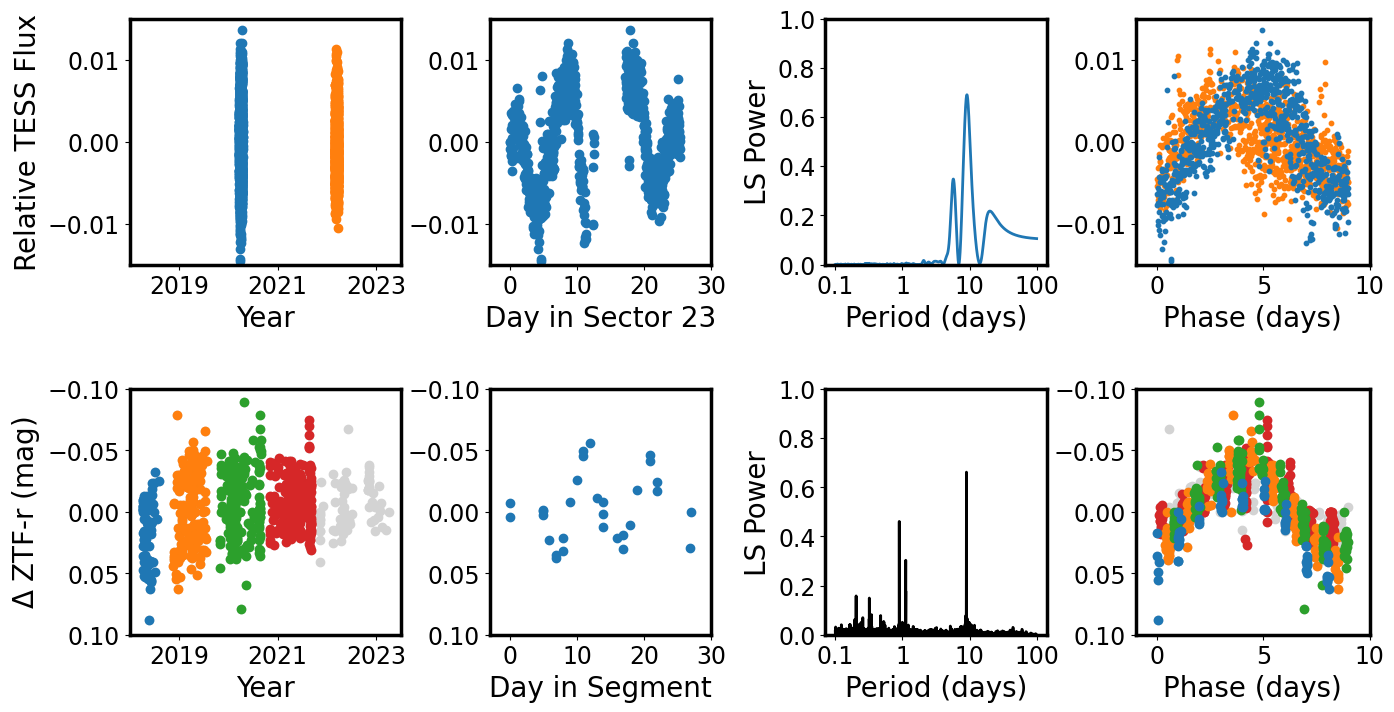}}
\vspace{-0.2cm}
\caption{Period analysis for \object{Gaia DR3 3957357582063045120}. Panels on the top row show TESS data (Cycles 23 and 49); those on the bottom row show ZTF data. The leftmost panels show all available photometric data for both surveys (at the time of our analysis), highlighting the years-long coverage provided by ZTF. The center-left panels show representative segments of length equal to a single TESS sector ($\approx$27~days), highlighting the relative sparseness of the ZTF light curves. The center-right panels show Lomb--Scargle periodograms for TESS Sector 23 and for the full ZTF light curve. The rightmost panels present the phase-folded light curves. Despite differences in data quality (total baseline, duration of each observing sector/season, cadence, precision, angular resolution), we obtain a \Prot\ = 9.17 d for this M3 stars from both datasets.} 
\label{fig_lc_example}
\end{figure*}

\begin{figure}[!h]
\includegraphics[width=1\columnwidth]{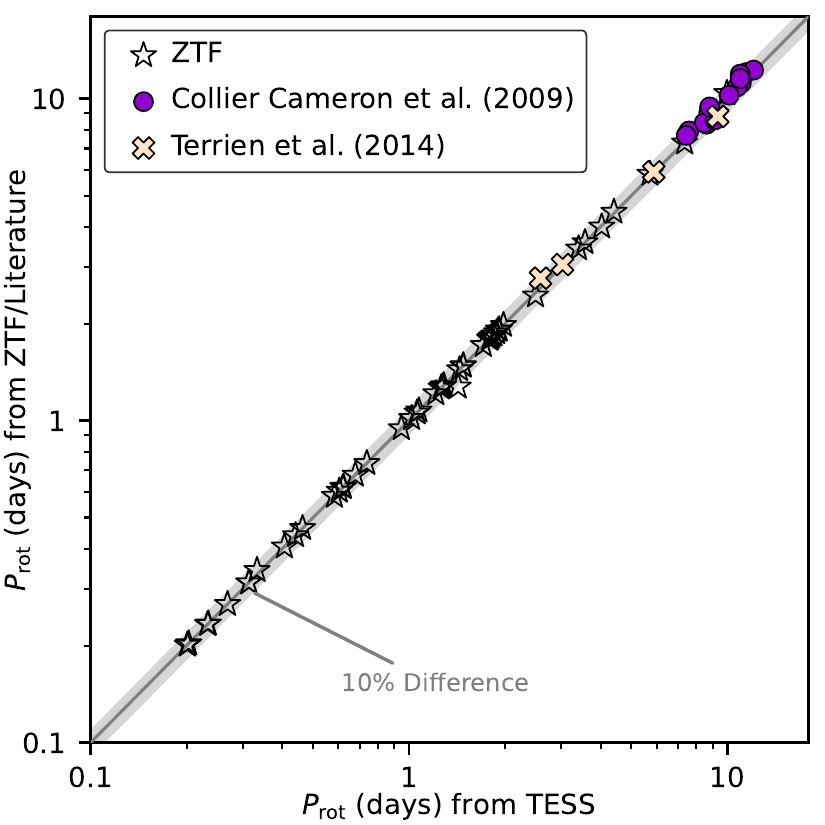}
\caption{\Prot\ from TESS vs.~\Prot\ from ZTF (black stars), \citet[][purple circles]{collier2009}, and \citet[][beige crosses]{terrien2014} for the 66 Coma Ber stars with more than one \Prot\ measurement. The solid gray line is the 1:1 match, and the shaded area the $\pm$10\% difference range. The overall agreement between the measurements is excellent, with a median difference $<$1\%.}
\label{fig:Prot_comp}
\end{figure}

\section{Rotation Periods}\label{sec:rot}
\subsection{Revisiting literature periods}

The two primary sources for Coma Ber \Prot\ measurements are the surveys of \citet{collier2009} and of \citet{terrien2014}. Nineteen \citet{collier2009} \Prot\ values are included in our rotation catalog for the cluster once our membership criteria are applied. We also include seven \Prot\ values measured by \cite{terrien2014}. Two stars have \Prot\ measurements in both of these surveys.

\subsection{Measuring new periods...}
\subsubsection{from TESS light curves}

The Transiting Exoplanet Survey Satellite \citep[][]{ricker2015} surveyed members of Coma Ber in Sectors 15 and 16 (2019 Aug 15 to Oct 06), 22 to 24 (2020 Feb 19 to May 12), 45 and 46 (2021 Nov 06 to 2022 Jan 28), and 49 and 50 (2022 Feb 26 to Apr 22). We downloaded 40$\times$40 pixel cutouts from the Full-Frame Images \citep{TESSdata} with {\tt TESScut} \citep{TESScut}, then produced photometric light curves using Causal Pixel Modeling implemented in the {\tt unpopular} Python package \citep{Hattori2022}. 

We then measured \Prot\ from Lomb-Scargle periodograms using an interactive program that allowed us to select light curves from each sector and visually validate periods. Given the relatively short duration of the observations of each sector ($\approx$27 d), and the large gap in the light curves midway through the observations due to the data downlink, our search was restricted to \Prot~$\lapprox$~15~d. For stars observed more than once by TESS, we used the light curves from the different sectors as a way of testing the robustness of our measurements.\footnote{Ideally one would  stitch together light curves from multiple TESS sectors to measure longer \Prot\ for a star observed more than once. In practice, however, this has proved challenging  \citep[e.g.,][]{anthony2022}. New approaches involving machine learning may eventually allow us to extend our investigation to longer \Prot\ \citep[cf.][]{claytor2024}.} 

The top row of Figure~\ref{fig_lc_example} illustrates our analysis for \object{Gaia DR3 3957357582063045120}, an M3 member for which we measured \Prot\ = 9.17~d. In total, we measured \Prot\ values for 128 Coma Ber members based on their TESS light curves. 

\subsubsection{from ZTF light curves}
The Zwicky Transient Facility \citep[][]{masci2018} has observed Coma Ber regularly since 2018 Mar 25. Following \citet{curtis2020}, we extracted differential photometry for each of our target stars from the archival ZTF imaging \citep{ZTF_data} using nearby reference stars in the field. As is evident in the bottom row of Figure~\ref{fig_lc_example}, which shows our analysis of the ZTF data for the same M3 member, the resulting light curves are much sparser and more irregularly sampled than those from TESS. Nevertheless, the ZTF-derived \Prot\ can provide confirmation of the TESS-derived periods in cases where a star was observed from the ground and from space. And because the ZTF light curves cover much longer timespans (close to 2000 d in this case), they can also be used to measure periods for rotators with \Prot $>$ 15 d, the \emph{de facto} limit for confident TESS-derived periods.

In total, we measured \Prot\ values for 75 Coma Ber members based on their ZTF light curves.  

\medskip
Figure \ref{fig:Prot_comp} compares our TESS \Prot\ values and the \Prot\ values for the 66 stars that also have either a measurement from ZTF data (48 stars) or from \citet{collier2009} and \citet{terrien2014} (19 stars, including one with a ZTF-derived \Prot). The agreement between these \Prot\ measurements is excellent: the median difference is $<$1\%. Our approach is therefore simply to average the different \Prot\ for these stars, and to use this averaged value for the rest of our analysis.

Combining the literature and our new TESS and ZTF measurements, we obtained a sample of 161 stars with a \Prot. Of these \Prot, all but 24 are new measurements.\footnote{Of the 24 literature periods, the TESS data allowed us to recover 15 from \citet{collier2009} and five from \citet{terrien2014}; one of those stars is in both surveys.}

\begin{figure*}
\centerline{\includegraphics{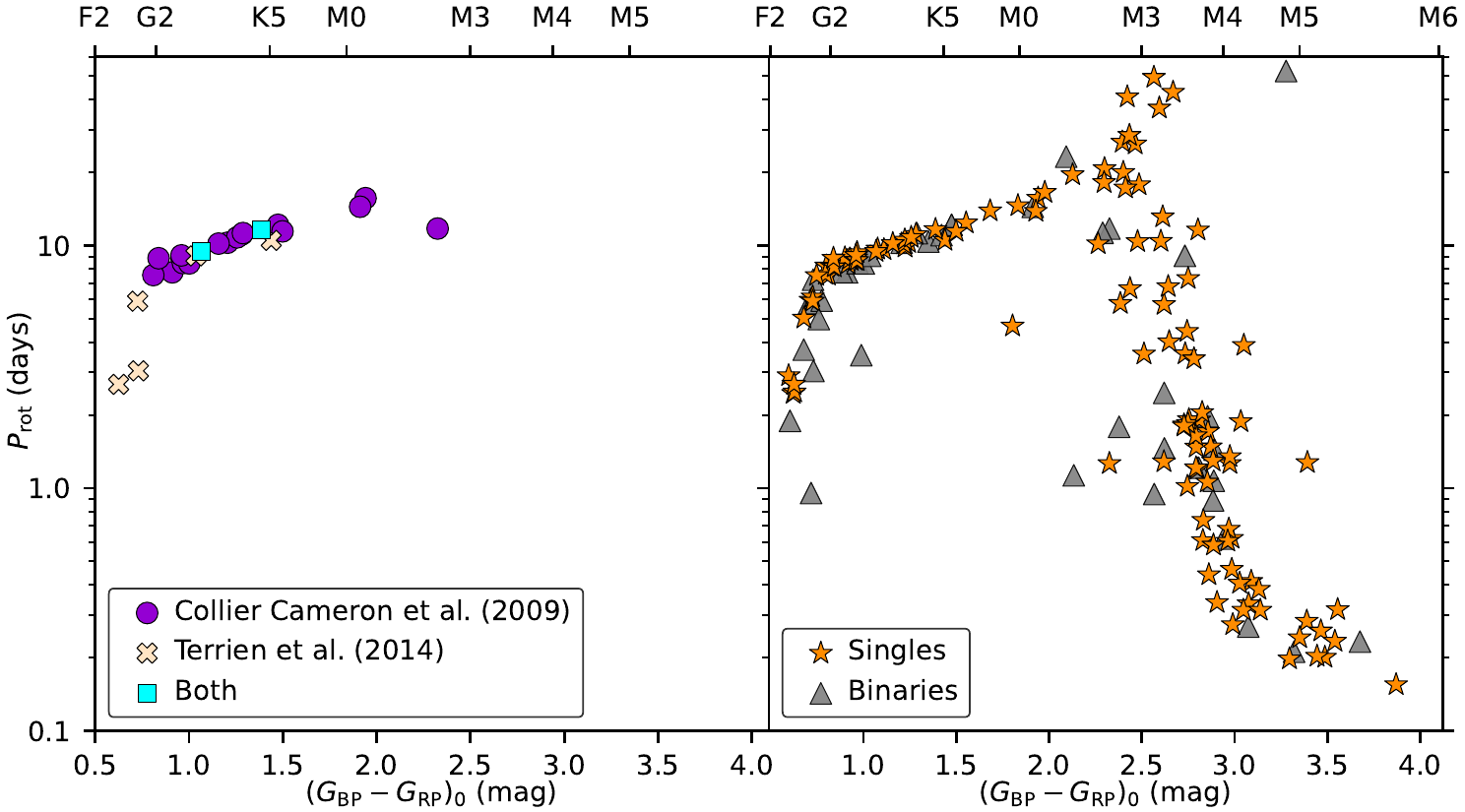}}
\caption{Color--period diagram for Coma Ber members before (left panel) and after (right) our TESS and ZTF measurements. In the left panel, the purple circles are stars with \Prot\ measured by \citet{collier2009}, the beige crosses are \Prot\ from \citet{terrien2014}, and the two cyan squares are stars with periods published by both sets of authors. There are 24 stars on this plot, mostly G and K stars. In the right panel, we show the updated \Prot\ distribution for the cluster, which now includes 161 stars. We highlight single members with orange stars, and candidate/confirmed binaries with gray triangles.}
\label{fig:CPD}
\end{figure*}

Figure~\ref{fig:CPD} shows the color--period diagram before (left panel) and after (right panel) the addition of our new TESS and ZTF \Prot\ measurements. The 161 \Prot\ values are included our Coma Ber membership catalog, described in Table~\ref{tab:cat}.

\subsection{Calculating Rossby numbers}\label{sec:Rossby}
For the past four decades at least, studies of the rotation--activity relation have chosen to quantify rotation by using the Rossby number \Ro, defined as \Prot/$\tau$, where $\tau$ is the convective turnover time, rather than \Prot\ \citep[e.g.,][]{Noyes1984}. As $M_\star$ decreases, the depth of the convective envelope increases as a fraction of the stellar radius, so that \Ro\ is a mass-independent parametrization of the star's rotation. 

To obtain $\tau$, we used the empirical \citet{Wright2018} $M_\star$--log($\tau$) relation, which is based on period and X-ray luminosity measurements for almost 850 stars ranging from 0.08 to 1.36~\Msun. With these $\tau$ values, we calculated \Ro\ for the 161 cluster stars with a measured \Prot. Our \Ro\ values range from $0.0012$ to $0.77$, comparable to the range for our samples of rotators in Praesepe and the Hyades \citep{Nunez2024}. We include these $\tau$ and \Ro\ values in our membership catalog, described in Table~\ref{tab:cat}.

\section{Optical Spectroscopy}\label{sec:spec}
\subsection{From our MDM observations}\label{MDM}

We obtained 214 spectra for 154 Coma Ber members over the course of 16 runs on the Hiltner 2.4-m telescope at the MDM Observatory, Kitt Peak, AZ. Our survey took place over 13 years (2011 Feb to 2024 May), and the instrumental set-up for our observations changed in 2018 and again in 2019 (see Table~\ref{table:MDM}). 

During our first five runs, we used the MDM Observatory Modular Spectrograph (ModSpec) with the Echelle detector, configured to cover the wavelength range from 4500 to 7500 \AA\ with 2.0~\AA\ pixel$^{-1}$ and spectral resolution $R \approx 1700$.

We then began to use the Ohio State Multi-Object Spectrograph (OSMOS). In 2018 Feb, during our first OSMOS run, we used the red 4K detector (R4K) with the VPH red grism, an OG-530 longpass filter, and 3$\farcs$0 center slit. This provided coverage from 6000 to 10,000~\AA\ (with peak efficiency near 7100 \AA), $\approx$0.7~\AA\ pixel$^{-1}$, and $R \approx 1600$.

We then switched to using the blue 4K detector (4K) with the VPH blue grism and a 1$\farcs$2 inner slit. This provided coverage from 4000 to 6800 \AA\ (with peak efficiency at 6400 \AA), 0.7\AA\ pixel$^{-1}$, and $R \approx 1600$. The majority of our spectra were obtained in this configuration. 

Spectra obtained before 2021 were processed with a PyRAF\footnote{\url{https://pypi.org/project/pyraf/}} script. Spectra obtained since 2021 were processed with {\tt PypeIt} \citep[Version $\geq$1.10.1.dev3+g52d10edd;][]{Prochaska2020, Prochaska2020zndo}. The PyRAF script and {\tt PypeIt} performed standard tasks, including trimming, overscan- and bias-correcting, cosmic-ray cleaning, flat-fielding, extracting, and dispersion-correcting and flux-calibrating the spectra. 

\begin{figure*}[t]
\centerline{\includegraphics{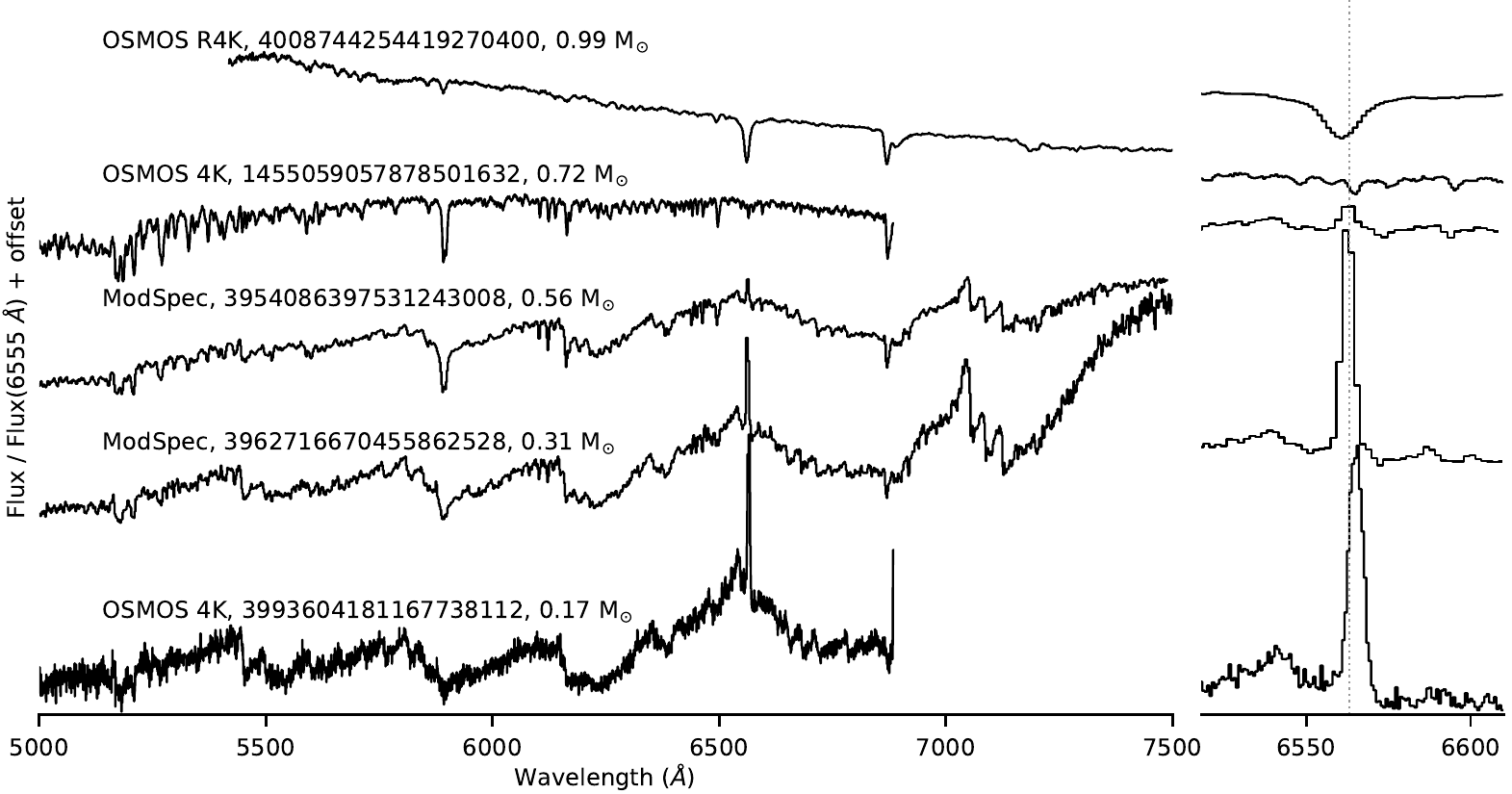}}
\vspace{-0.2cm}
\caption{Five representative spectra obtained with the Hiltner 2.4-m telescope at MDM Observatory, each labeled with the instrument used and the star's DR3 identifier and our estimate of its mass. The spectra are normalized to the flux at 6555~\AA\ and offset for plotting purposes. The right panel shows close-ups of the \Ha\ line; the dotted line is at 6563 \AA, the position of the line core. These spectra illustrate the change from \Ha\ absorption to strong emission as we go from solar-like to late-M stars.}
\label{fig:sample_spectra}
\end{figure*}

Figure~\ref{fig:sample_spectra} shows five example spectra, illustrating the varying strength of the \Ha\ line in our sample. The median signal-to-noise ratio ($S/N$) at the H$\alpha$ line core (6563 \AA) was $\approx$75. The reduced MDM spectra are available online.\footnote{Available from the Columbia University Academic Commons under a CC0 license: \url{ https://academiccommons.columbia.edu/doi/10.7916/p1ta-qn53}.\label{fn:commons}} 

\begin{deluxetable}{@{}llc@{}}
\centering 
\tabletypesize{\footnotesize}

\tablecaption{MDM observations of Coma Ber
 \label{table:MDM}}

\tablehead{
\colhead{Date} &
\colhead{Instrument/} &
\colhead{\# of}\\[-0.1in]
\colhead{} &
\colhead{Detector} &
\colhead{Spectra}
}

\startdata
2011 Feb 08$-$Feb 10 & ModSpec / Echelle & 14 \\
2012 Feb 17$-$Feb 20 & ModSpec / Echelle & 9 \\
2015 Feb 20$-$Feb 21 & ModSpec / Echelle & 17 \\
2016 Jan 29$-$Feb 03 & ModSpec / Echelle & 23 \\
2017 Feb 15$-$Feb 20 & ModSpec / Echelle & 11 \\
2018 Feb 06$-$Feb 10 & OSMOS / R4K & 5 \\
2019 Feb 28$-$Mar 03 & OSMOS / 4K & 23 \\
2019 Apr 10$-$May 07 & OSMOS / 4K & 23 \\
2021 Apr 16$-$Apr 19 & OSMOS / 4K & 13 \\
2021 May 04$-$May 07 & OSMOS / 4K & 12 \\
2022 Apr 13$-$Apr 23 & OSMOS / 4K & 40 \\
2023 Mar 29$-$Mar 31 & OSMOS / 4K & 4 \\
2024 Mar 01$-$Mar 07 & OSMOS / 4K & 5 \\
2024 Mar 21$-$Mar 29 & OSMOS / 4K & 6 \\
2024 Apr 20$-$Apr 21 & OSMOS / 4K & 4 \\
2024 May 08          & OSMOS / 4K & 5 \\
\enddata

\vspace{-.2cm}

\end{deluxetable}

\subsection{From the LAMOST and SDSS archives}\label{lamost}

We supplemented our spectroscopic survey with spectra from the LAMOST Data Release 10 (DR10) low-resolution catalog.\footnote{\url{https://www.lamost.org/dr10/v2.0/}} These spectra are flux- and wavelength-calibrated and sky-subtracted, and cover 3690 to 9100~\AA\ with $R=$~1800 at 5500~\AA. After excluding spectra with a $S/N \lapprox$~5, we found 299 spectra for 97 Coma Ber stars. 

We also searched for spectra in the SDSS Science Archive Server.\footnote{\url{https://dr18.sdss.org/home}; last consulted 2025 Apr 1.} SDSS spectra are sky-subtracted, corrected for telluric absorption, spectrophotometrically calibrated, and calibrated to heliocentric vacuum wavelengths. The wavelength coverage is 3800 to 9200 \AA\ with $R$ $\approx$ 2000 at \halpha. After excluding spectra with $S/N$ $\lapprox$~5, we found 13 SDSS spectra for nine Coma Ber members.

\begin{figure}
\includegraphics{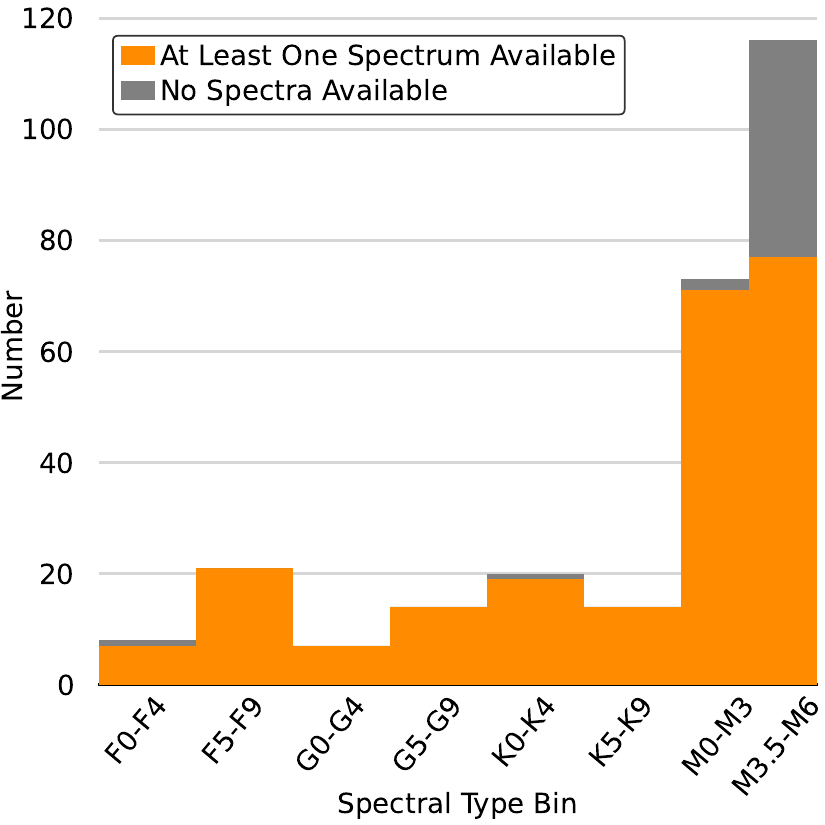}
\caption{Stacked histogram illustrating our excellent optical spectral coverage for Coma Ber. The orange bars correspond to the stars for which we have at least one good quality optical spectrum with which to measure the \halpha\ EW (246 of 291 cluster members). Gray bars indicate the members without spectra.}
\label{fig:cov}
\end{figure}

Combining our MDM spectra with those from the LAMOST and SDSS archives, we have at least one spectrum for 245 of the 291 high-confidence and candidate members of Coma Ber. This represents exceptional spectral coverage (85\%), as illustrated by the histogram in Figure~\ref{fig:cov}. Twenty-five percent of our spectroscopic targets are known or candidate binaries.

\begin{figure*}[!t]
\includegraphics{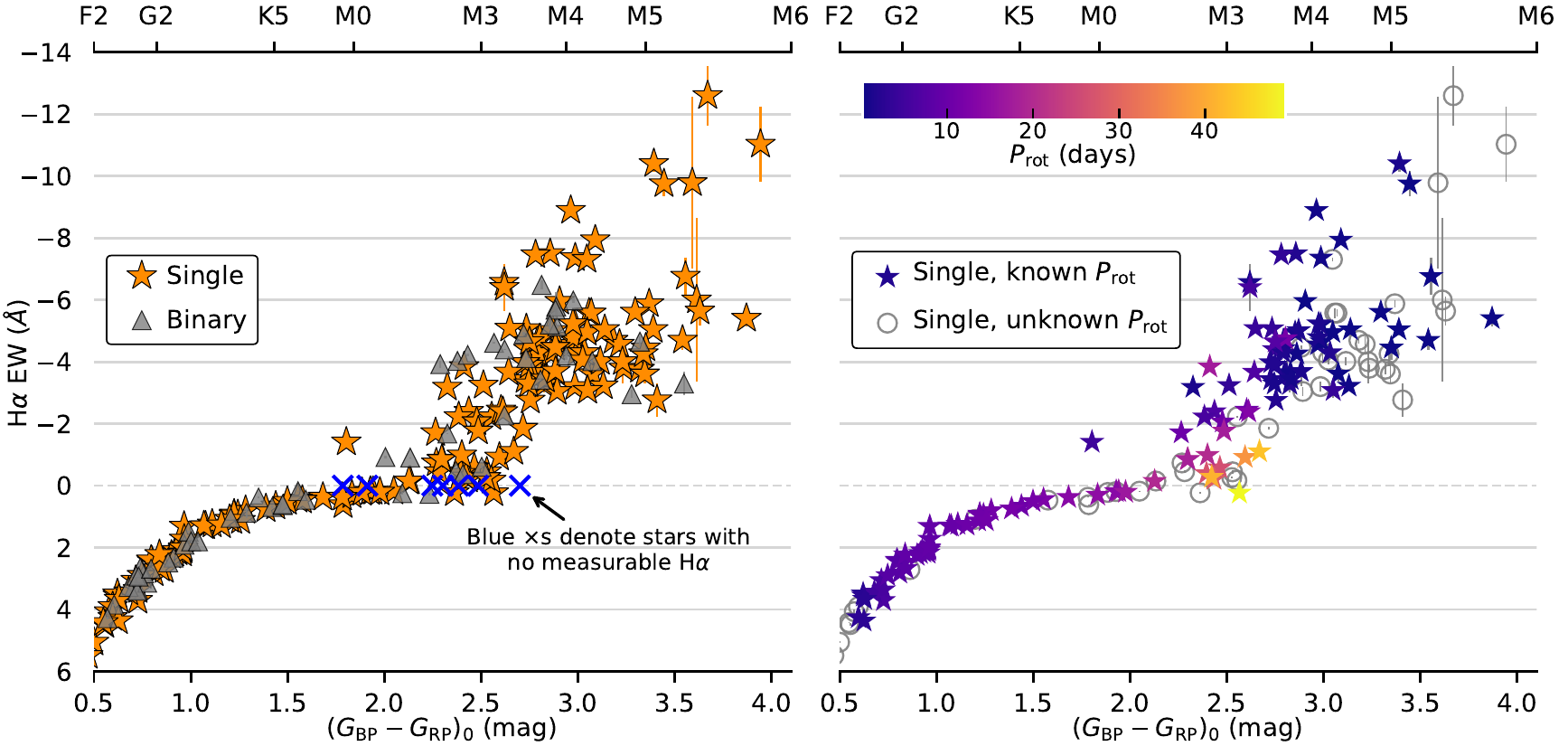}
\caption{Measured \Ha\ EW as a function of \bpminusrp\ for low-mass stars in Coma Ber. Negative EW values correspond to emission. The left panel distinguishes between likely single stars (orange stars) and known and candidate binaries (gray triangles). Blue $\times$ symbols denote stars for which we could not measure an EW and to which we assigned EW = 0~\AA. The right panel distinguishes between single members with known \Prot\ (colored stars, following the inset color bar) and single members without a measured \Prot\ (gray circles). For clarity, we excluded from the panels one outlier star with \bpminusrp\ = 2.8 mag and EW = $-76$ \AA. We also excluded stars with \bpminusrp\ $<$ 0.5 (spectral types earlier than $\approx$F5), as they are not relevant to our analysis.}
\label{fig:EW_color}
\end{figure*}

\subsection{Measuring equivalent widths}
We used \texttt{PHEW} \citep{PHEW} to measure \halpha\ equivalent widths (EWs) and 1$\sigma$ uncertainties for our spectra. \texttt{PHEW}, which incorporates PySpecKit \citep[][]{Ginsburg2011} and a Monte Carlo iterator, is described in detail in \citet{Nunez2024}. 

The primary difference between \texttt{PHEW} and a standard EW-measurement tool is how it estimates uncertainties: the code performs 1000 Monte Carlo iterations by re-sampling the flux measurements within the flux uncertainties, or, if these uncertainties are unavailable, by adding Gaussian noise to the flux spectrum. It then calculates the standard deviation of the 1000 EWs, which we adopt as the 1$\sigma$ EW uncertainty. 

For the 129 stars for which we have more than one spectrum, we calculated for each star the error-weighted mean EW (and the weighted mean 1$\sigma$ uncertainty) and adopted those as the EW values and 1$\sigma$ uncertainties, respectively, for the star. These measurements are included in our  membership catalog, described in Table~\ref{tab:cat}.

For two early-to-mid M dwarfs (\object{Gaia DR3 3952459012947536256} and \object[Gaia DR3 4009049575054518400]{4009049575054518400}), we had one spectrum with a significantly larger ($>$50\%) EW than that measured from the other spectrum/a. Such a large deviation in the EW is likely due to an unusual event, such as a flare, and as such that measurement is not representative of the baseline magnetic activity in those stars. We therefore clipped the outlier EW measurements when calculating the final mean EW values for these stars.

Figure~\ref{fig:EW_color} shows our measured EW values as a function of \bpminusrp\ for high-confidence and candidate members of Coma Ber. The left panel distinguishes likely single members from known and candidate binary members. We highlight seven stars for which we found no measurable \halpha; we assigned these stars EW = 0~\AA. The right panel only includes likely single stars; those with measured periods are color-coded by the \Prot\ value.

To better visualize the dependence of the EW on color, we omitted from Figure~\ref{fig:EW_color} one Coma Ber outlier (\object{Gaia DR3 3923649162599782784}) with \bpminusrp\ = 2.8 mag ($\approx$M4 spectral type) and EW~=~$-76$ \AA, as this star was likely flaring when it was observed. We also excluded stars with \bpminusrp\ $<$ 0.5 mag (spectral types earlier than $\approx$F5), as their lack of significant convective envelopes implies a different rotational evolution than the Sun-like stars we focus on. 

\subsection{Correcting for photospheric H$\alpha$ absorption}\label{sec_relativeEW}
The \halpha\ line observed in magnetically active stars is a combination of emission in the chromosphere and absorption in the photosphere \citep{Stauffer1986}. Only the former is primarily a consequence of magnetic heating. As in \citet{Nunez2024}, we therefore used the empirical model of \citet{Newton2017} to estimate the EWs of the (quiescent) photospheric \halpha\ absorption lines, whose strength is a function of mass.\footnote{The \citet{Newton2017} model is for stars with $M_\star$ between 0.1 and 0.8~\Msun. In our Coma Ber sample, all stars with \halpha\ emission are within this mass range.} 

We then determined relative EWs, which reflect only the chromospheric contribution to our \halpha\ measurements, by subtracting the quiescent EWs from our measured EWs. We include these relative EWs in our membership catalog and used them to calculate the ratio of the \halpha\ luminosity to the bolometric luminosity, \LLH, as described below. 

\subsection{Calculating the $\chi$ factor for stars with measured \halpha\ equivalent widths}\label{sec:chi}

When trying to establish relative levels of magnetic activity, using the ratio of the \halpha\ luminosity to the bolometric luminosity, \LLH, rather than just EWs, is more meaningful for samples of stars ranging significantly in mass. For spectra that are not well flux-calibrated, however, obtaining reliable values for the continuum flux near the \halpha\ line can be difficult, and so calculating \Lha\ directly is not always possible. Instead, one can use the $\chi$ factor, a tabulated set of \halpha-continuum-to-bolometric flux ratios for stars varying in color/\teff, to convert EWs to \LLH\ \citep[\LLH\ = $-$EW~$\times\ \chi$;][]{walkowicz2004}.

\citet{Douglas2014} presented an empirical relation between \teff\ and $\chi$ based on PHOENIX ACES model spectra \citep{Husser2013} for stars with \teff\ from 2500--5200 K, and \citet{Nunez2024} expanded the relation to cover 2300 to 6500 K. We used this expanded relation to calculate $\chi$. First, we derived \teff\ using an empirical $M_K - \teff$ relation.\footnote{Derived from the table of E. Mamajek, version 2022.04.016. See Footnote~\ref{note}.} We linearly interpolated between the $M_K$ values in the empirical relation to obtain \teff. Next, we linearly interpolated between the \teff\ values in the $\teff-\chi$ relation to obtain $\chi$ and thus \LLH. 

We include the \teff, $\chi$, and \LLH\ values in our membership catalog, described in Table~\ref{tab:cat}.

\section{X-Ray Data} \label{sec:xray}

\subsection{Matching to serendipitous source catalogs}

Coma Ber's proximity and resulting large angular size make it a poor target for pointed X-ray observations. But its large footprint on the sky means that some of its members were likely detected during  observations of unrelated targets. We followed the approach described in  \citet{Nunez2022} to search the serendipitous X-ray source catalogs for matches to cluster members. 

We found counterparts to five stars in the Chandra X-ray Observatory (Chandra) Source Catalog \citep[CSC 2.1,][]{CSC2} using a 15\asec\ matching radius;\footnote{The Chandra on-axis angular resolution is sub-arcsecond, but it quickly degrades towards the edges, reaching up to $\approx$15\asec\ half-energy width at 1.5 keV $\approx$10\amin\ from the aimpoint.} the largest separation between a star and its X-ray counterpart was 5$\farcs$3. For these sources, the instrumental count rate is in the 0.5--7.0 keV band.

We found counterparts to two stars in the Neil Gehrels Swift Observatory (Swift) XRT Point Source Catalog \citep[2SXPS,][]{2SXPS} using a 15\asec\ matching radius; the largest separation was 8$\farcs$5.\footnote{Swift's XRT on-axis angular resolution is between 15 and 18\asec\ half-energy width at 1.5 keV.} The count rate is in the 0.3--10.0 keV band.

We found counterparts to eight stars in the X-ray Multi-Mirror Mission Newton (XMM) Serendipitous Source Catalog \citep[4XMM-DR13,][]{4XMM} using a 15\asec\ matching radius;\footnote{XMM's EPIC on-axis angular resolution is $\approx$15\asec\ half-energy width at 1.5 keV.} the largest separation was 2$\farcs$7. The count rate is in the 0.2--12.0 keV band.

For each match, we recorded the detection likelihoods and variability flags, when available. When more than one X-ray source matched to a Coma Ber star, we assumed the closest source to be the star's counterpart. 

\subsection{Matching to all-sky survey data}

We also searched for detections of Coma Ber stars in the all-sky X-ray surveys conducted with the R\"ontgen Satellite (ROSAT) and, most recently, as part of the extended ROentgen Survey with an Imaging Telescope Array  \citep[eROSITA;][]{erosita}. For ROSAT, we again followed the approach described in \cite{Nunez2022} to identify counterparts to our stars in the Second ROSAT Source Catalog of Pointed Observations with the Position Sensitive Proportional Counters (PSPCs) catalog \citep[2RXP,][]{2RXP}, the Second ROSAT All-Sky survey catalog \citep[2RXS,][]{2RXS}, and the ROSAT High-Resolution Imager Pointed Observations catalog \citep[1RXH,][]{1RXH}. 

Using a 30\asec\ matching radius,\footnote{ROSAT's PSPC on-axis angular resolution is $\approx$45\asec\ half-energy width at 1.5 keV.} we found 18 2RXP and 16 2RXS counterparts to Coma Ber members. The largest separation between a star and its ROSAT counterpart was 26$\farcs$4. The 2RXP and 2RXS count rates are in the 0.11--2.02 keV band.

The first data release of the eROSITA All-Sky survey catalog\footnote{\url{https://erosita.mpe.mpg.de/dr1/AllSkySurveyData_dr1/Catalogues_dr1/}} \citep[eRASS DR1 v1.1,][]{eRASS} covers Galactic longitudes between 180 and 360\arcdeg. We searched for counterparts in two of the eRASS catalogs. The Main catalog includes sources detected in Band 0, corresponding to 0.2--2.3 keV, with a detection likelihood $\geq$6. The Supplementary catalog includes sources detected in the same band, but with a detection likelihood between 5 and 6, and is therefore expected to include more spurious sources. 

Using a 30\asec\ matching radius,\footnote{The eROSITA on-axis angular resolution is 16$\farcs$1 half-energy width at 1.5 keV.} we found X-ray counterparts for 76 Coma Ber members in the Main catalog; the largest separation between a star and its X-ray counterpart was 18$\farcs$5, and the median, 4$\farcs$7. These X-ray sources were offset by $\lapprox$3$\times$ the positional uncertainty of the eRASS source. 

Three of the matches were with X-ray sources that had no count rate uncertainties. Two additional matches were with X-ray sources with an extent likelihood $>$0.\footnote{The extent likelihood quantifies the probability that an X-ray source is spatially extended by comparing the results of fitting detections with both a point-source and an extended-source model.} We excluded these five matches from our analysis.

We found five X-ray counterparts in the Supplementary catalog; the largest separation was 13$\farcs$1, and the median, 5$\farcs$1. These X-ray sources were offset by $\lapprox$2.5$\times$ the positional uncertainty of the eRASS source. We inspected each source individually and found no evidence that they were spurious. In fact, four had a detection likelihood $>$6 in the P2 band (0.5--1.0 keV). 

After consolidating our findings in the serendipitous catalogs and all sky surveys, we have 96 Coma Ber members with at least one X-ray detection; 26 have two or more. Table~\ref{tbl:Xcols} describes the columns in our catalog of these X-ray sources, available online.

\begin{deluxetable}{@{}ll}
\tabletypesize{\scriptsize} 

\tablecaption{Columns in the X-ray source catalog \label{tbl:Xcols}}

\tablehead{
\colhead{Column} & \colhead{Description} 
}

\startdata
1      & Provenance of X-ray data\tablenotemark{a}\\
2      & IAU Name \\
3      & Instrument \\
4, 5   & R.A., Decl.~for epoch J2000\\
6      & X-ray positional uncertainty \\
7      & Detection likelihood $L$ \\
8      & Net counts \\
9, 10  & Net count rate and 1$\sigma$ uncertainty \\
11     & Energy band used \\
12     & Exposure time \\
13     & Variability flag: (0) no evidence for variability; \\
       & (1) possibly variable; (2) definitely variable \\
14, 15 & Unabsorbed energy flux and 1$\sigma$ uncertainty in the \\
       & 0.1--2.4 keV band \\
16     & Gaia DR3 designation of the optical counterpart \\
17     & Separation between X-ray source and optical counterpart \\
\enddata

\tablenotetext{a}{2RXP; 2RXS; 2SXPS; 4XMM; CSC; eRASS Main; eRASS Supplement.}

\vspace{-.2in}

\end{deluxetable}

\subsection{Calculating X-ray fluxes and luminosities}\label{sec:fx}

We homogenized our X-ray data following the approach described in \citet{Nunez2022}. Briefly: we converted instrumental count rates into unabsorbed energy fluxes, \fx, in the 0.1$-$2.4 keV energy band\footnote{This corresponds roughly to the ROSAT energy band, and has historically been used for studies of stellar coronae.} using energy conversion factors derived with WebPIMMS.\footnote{\url{https://heasarc.gsfc.nasa.gov/cgi-bin/Tools/w3pimms/w3pimms.pl}}

We used a one-temperature thermal APEC model \citep{Smith2001} with a metal abundance of 0.2 or 0.4, depending on whether the optical counterpart had \gminusk\ $>$ 1.8 or $\leq$1.8 mag. We assumed a plasma temperature of 0.6845 keV, and a negligible extinction by neutral atomic hydrogen $N_\mathrm{H}\ \lapprox\ 10^{18}$ cm$^{-2}$. No X-ray detection had a sufficient numbers of counts for us to perform a meaningful spectral analysis. The catalog described in Table~\ref{tbl:Xcols} includes the derived \fx\ for each X-ray source.

\begin{figure*}[t!]
\centerline{\includegraphics{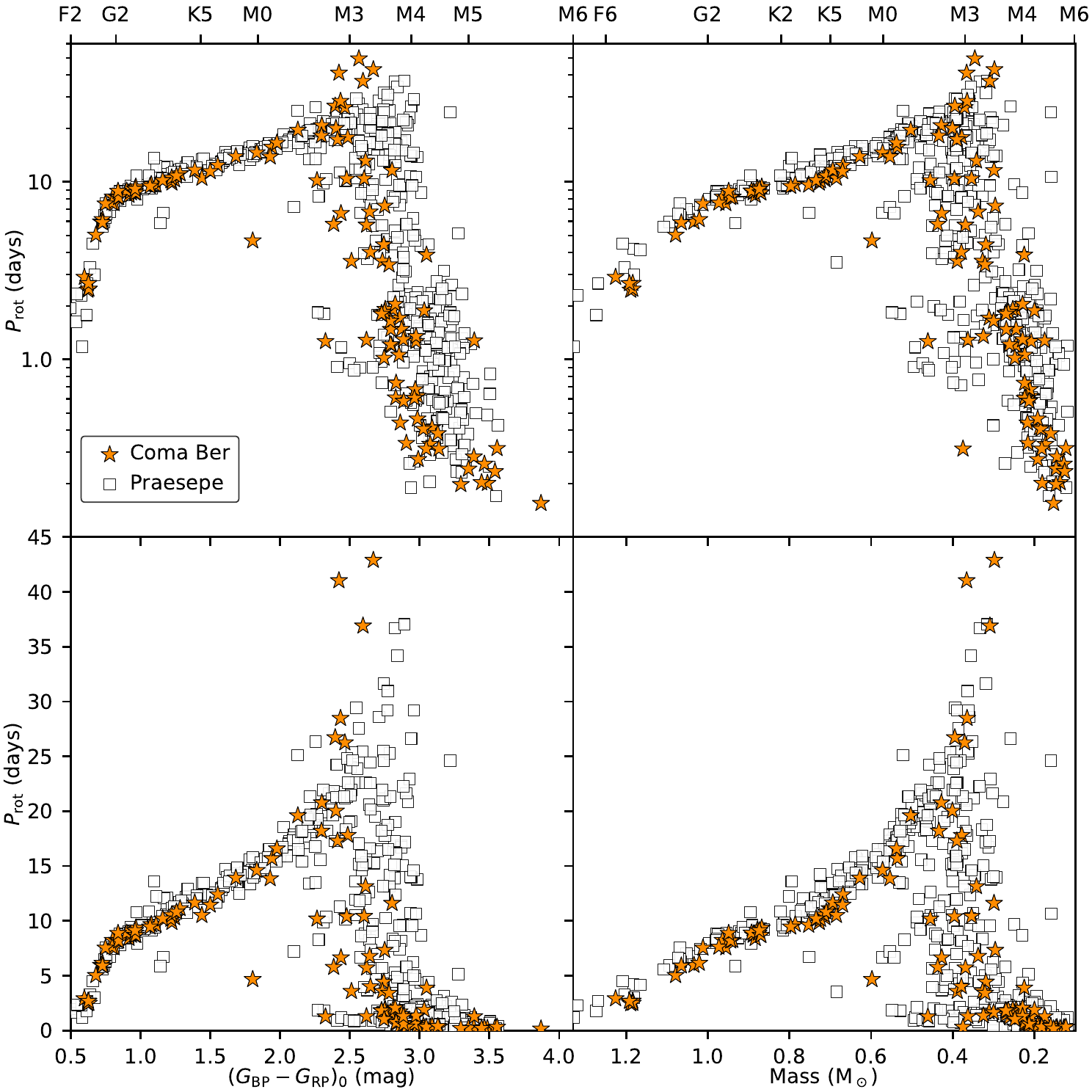}}
\caption{Color--period (left column) and mass--period (right column) diagrams for likely single members of Coma Ber (orange stars) and Praesepe \citep[white squares; data from][]{Nunez2024}. In each column, the bottom and top panels show the same data; we plot the \Prot\ linearly at the bottom and logarithmically on the top. The agreement in the \Prot\ distributions for the hotter  (F through early M) stars in our samples is very clear in both panels on the left; the offset in the distributions for the mid and late M dwarfs is most obvious in the top left panel. To produce the two panels in the right column, we have transformed the Gaia colors into masses. The M dwarf \Prot\ distributions now agree (as is most evident in the top right panel), whereas the distributions for the $\approx$solar-mass stars are offset, as is most clearly visible for the 0.6--0.9 \Msun\ stars in the bottom right panel.}
\label{fig:CPDPraesepe}
\end{figure*}

For stars with more than one detection, we calculated the error-weighted average of the \fx\ values and adopted it as the \fx\ for that star. We then calculated \lx\ using the adopted distances to each star. 
The \lx\ values for the stars with X-ray detections are included in our membership catalog, described in Table~\ref{tab:cat}. 

\section{Discussion}\label{sec:disc}

\subsection{Rotation period distribution}\label{sec:rot_dist}

Metallicity is expected to influence the rotational evolution of Sun-like stars. For a given mass, more metal-rich stars will have convective envelopes that extend deeper into the stellar interiors \citep[e.g.,][]{vansaders2013}.\footnote{For a given mass and \Prot, this implies that these more metal-rich stars will have smaller Rossby numbers.} Theoretical work indicates that this impacts the winds carried away from the stars' surfaces by their magnetic field lines, and that the torques experienced by stars over their lifetimes are therefore metallicity-dependent \citep[e.g.,][]{amard2020a}. 

\cite{amard2020b} found that, in a sample of $>$4000 0.85 to 1.3~\Msun\ stars observed with Kepler, the metal-rich ones rotate more slowly than the metal-poor ones, evidence for this metallicity-dependent magnetic braking, a result later confirmed by \citet{santos2023} and \citet{see2024}. Interestingly, however, not all studies have clearly detected this difference; see, e.g., \cite{avallone2022}.

While open clusters are an ideal laboratory to search for evidence of this metallicity dependence, we  have only a few examples of open clusters of the same age with metallicities that differ significantly (i.e., by as much as one or two tenths of a dex). The \Prot\ data for Coma Ber are therefore particularly valuable given the extensive existing data for the coeval Praesepe and Hyades.

The left column of Figure~\ref{fig:CPDPraesepe} shows the distribution of rotation periods as a function of Gaia color for likely single members of Coma Ber compared to that for Praesepe, as assembled by \cite{Nunez2024}. For clarity we plot the two distributions twice: once with linear \Prot\ (bottom panel), which allows for a closer examination of the behavior of slowly rotating late F, G, K, and early M stars in the two clusters, and once with logarithmic periods (top panel), which better illustrates the full range of \Prot\ for the mid and late M dwarfs. 

The color--\Prot\ relations for the late F, G, K, and early M stars are remarkably similar in the two clusters, despite the 0.2 dex difference in metallicity between them. However, the \Prot\ distributions appear to differ for stars later than $\approx$M3. While the range of measured periods for these stars is comparable, from $\approx$0.2 to 40 d, the distribution for Praesepe members is shifted slightly to redder colors relative to that for Coma Ber stars. (Equivalently, for a given \bpminusrp\ $\gapprox$ 2.5~mag, Praesepe stars are rotating more slowly than Coma Ber members.) 

We conducted 2D Kolmogorov–Smirnov (KS) tests to examine whether the two \Prot\ distributions are statistically distinct. When we considered the 77 stars with \bpminusrp\ $>$ 1.84~mag (spectral types later than M0) in Coma Ber with a measured \Prot\ and their 412 cousins in Praesepe, we found that p~=~0.002, well below the p~=~0.05 threshold that is generally used to indicate that one can reject the null hypothesis. In other words, the \Prot\ distributions for the coolest stars in the two clusters are indeed statistically different. 

A similar KS test comparing the Coma Ber distribution to that of the 130 M dwarfs in the Hyades with a measured \Prot\ \citep[again using the data of][]{Nunez2024} confirmed this result, with p = 0.004. While the Hyades is slightly older than Praesepe (by $\approx$60 Myr), the rotational distributions for the M dwarfs in the two clusters are indistinguishable \citep{douglas2019}. As a sanity check, and in light of the expanded list of rotators for both clusters published by \cite{Nunez2024}, we conducted the same 2D KS test between M dwarfs in these two clusters, and found that p~=~0.169, confirming that their \Prot\ are drawn from the same underlying distribution. 

We also calculated the KS D statistic, a measure of the maximum vertical distance between the empirical cumulative distribution functions of the two data sets. D values can be compared to reference D$_{\rm critical}$ values for a given ($1 - \alpha$) confidence level threshold; when D $>$ D$_{\rm critical}$, the null hypothesis can be rejected.

The D statistics confirm that the \Prot\ distributions are distinct, with values that are larger than D$_{\rm critical}$ for $\alpha = 0.05$: for the comparison between M dwarf rotators in Coma Ber and those in Praesepe, D~=~0.262, with D$_{\rm critical}$~=~0.184, and for these rotators in Coma Ber and in the Hyades, D~=~0.284 and D$_{\rm critical}$~=~0.213. Meanwhile, when comparing Praesepe and Hyades M dwarfs, D~=~0.127, which is smaller than D$_{\rm critical}$~=~0.149, confirming that the two sets of measurements appear to be drawn from the same underlying distribution. 

An initial reading of these results is that for F through early M stars, any observable difference in rotational evolution in color space due to a metallicity difference of 0.2 dex has been largely erased by an age of 700 Myr. By extension, one can apply a gyrochronological relation to these stars that uses their colors without needing to know their metallicities precisely, as their spin-down ages appear insensitive to a small variations in metallicity (up to at least 0.2 dex). 

By contrast, this metallicity difference does impact the observed spin-down evolution for the later M dwarfs. For example, an M3 star in Praesepe is more likely than its cousin in Coma Ber to have spun down sufficiently to be on the slow-rotating sequence defined in color space by more massive cluster members. Applying a gyrochronological relation to these stars without metallicity information is therefore likely to negatively impact estimates of their ages.  

This interpretation, while representative of what is happening in the observed (color) space, ignores the impact of metallicity on a star's color. For two stars of the same mass, a difference of 0.2 dex in metallicity should result in a difference of $\approx$0.1 mag in \bpminusrp, with the more metal-rich star being redder. An equivalent statement is that for this difference in metallicity, when considering two stars of the same color, the metal-rich one will be more massive by $\approx$10\%.

We plot in the right column of Figure~\ref{fig:CPDPraesepe} the same two \Prot\ distributions as in the right panel, but now as a function of mass rather than color. If we now consider stars with mass $\leq$0.6 \Msun\ (spectral types later than M0) separately from those with mass $>$0.6 \Msun, the difference with the results in color space is striking. The top right panel of Figure~\ref{fig:CPDPraesepe} shows that the \Prot\ distributions for the lowest-mass members of the two clusters are now well-matched; the offset seen in the top left panel for the mid and late M dwarfs is no longer visible. 

Another 2D KS test confirms this agreement in mass space of the \Prot\ distributions for the $\leq$0.6 \Msun\ stars. Comparing the periods for these stars in Coma Ber to those for their cousins in Praesepe, we find that p = 0.09, and D = 0.172 with D$_{\rm critical}$ = 0.181, implying that the \Prot\ are drawn from the same underlying distribution. 

Why do we no longer see an offset in the \Prot\ distributions for these mid and late M dwarfs? A plausible explanation is that, as explored in Section~\ref{sec:sat}, most of these stars are magnetically saturated, meaning that the strength of their magnetic field no longer depends on their rotation rate. When this is the case, differences in the depth of the convective envelope that result from differences in metallicity no longer result in differences in the torque applied on the stars by their winds, as these \emph{do not} depend on the Rossby number in this regime \citep[e.g.,][]{matt2015}. 

By contrast, if one focuses on the $>$0.6 \Msun\ stars in the bottom right panel of Figure~\ref{fig:CPDPraesepe}, there is a small offset between the two slow-rotating sequences that is most obvious for the K stars. These stars in Coma Ber have not spun down as much as their more metal-rich counterparts in Praesepe, consistent with the \cite{amard2020a} predictions   and as seen by \cite{amard2020b} for Kepler stars.  

\begin{figure}
\includegraphics{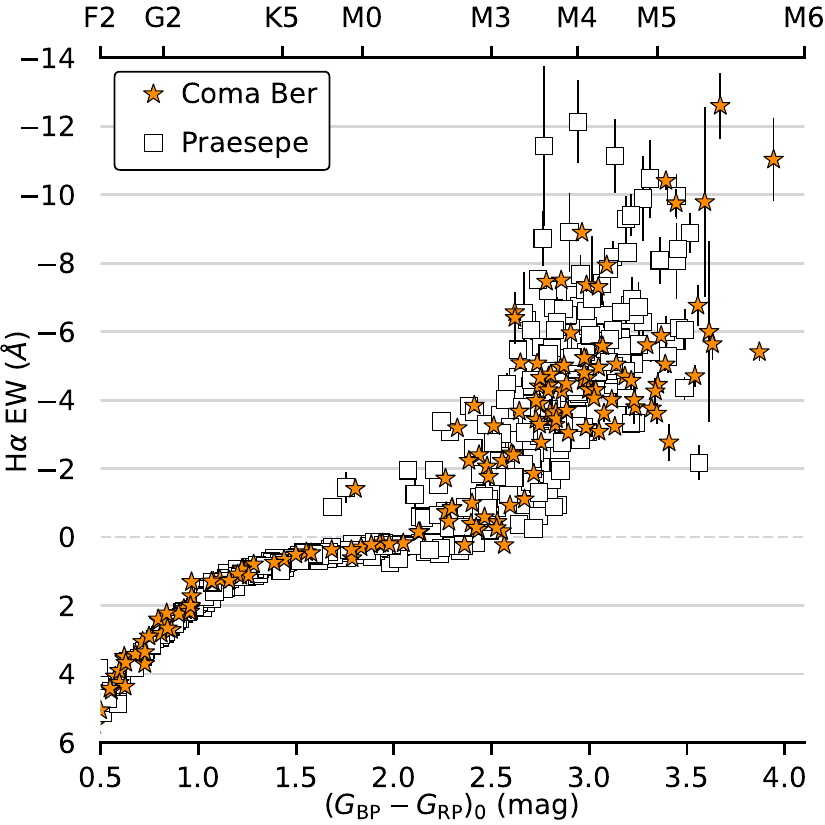}
\caption{Measured \halpha\ EW as a function of \bpminusrp\ for single members of Coma Ber (orange stars) and of Praesepe \citep[white squares; data from][]{Nunez2024}. We exclude the same outlier in Coma Ber as in Figure~\ref{fig:EW_color}. The distributions for the two clusters are remarkably similar, with stars earlier than $\approx$M2 being inactive, and later stars showing a comparable range of \halpha\ emission at a given color. Interestingly, the relative shift in the \Prot\ distribution for the later M dwarfs seen in the top left panel of  Figure~\ref{fig:CPDPraesepe} does not appear to have an equivalent in this space.
\label{fig:EW_color_Pr} }
\end{figure}

\subsection{\halpha\ equivalent width distribution}\label{disc:ha_dist}

Because of the underlying connection between a Sun-like star's angular-momentum content and its magnetic field generation \citep[e.g.,][]{Parker1993}, magnetic activity should also be a metallicity-dependent stellar property. But there are still relatively few large-scale observational studies of how metallicity impacts magnetic activity for stars of a given mass and \Prot. This is largely because of the difficulty involved in assembling large samples of  stars with \Prot\ measurements and measurements of proxies for magnetic activity, \emph{and} with a meaningful spread in metallicity.
 
Recent efforts based on Kepler observations have suggested that several manifestations of magnetic activity do correlate positively with metallicity \citep{see2021,see2023,santos2023,mathur2025}. For stars of a given mass rotating at a given \Prot, for example, the more metal-rich ones will exhibit more luminous flares \citep{see2023}. Other cases are more ambiguous, however; while \citet{santos2023} and \citet{mathur2025} found a (weak) correlation between metallicity and photometric variability, \citet{see2024}, who estimated ages for their sample, did not see evidence for such a correlation at a given mass and age.\footnote{It is worth remembering that photometric variability is a fairly indirect proxy for magnetic activity. As pointed out by \citet{see2024}, if two stars have the same level of activity but different inclinations, their light curves will show different levels of brightness variations, and their photometric activity indices will therefore differ as well.} By comparing our \halpha\ data for Coma Ber to the extensive data available for Praesepe and the Hyades, we can repeat the \citet{see2024} test for a more direct proxy for magnetic activity at $\approx$700 Myr. 

Figure~\ref{fig:EW_color_Pr} is a comparison of the distribution of our measured \halpha\ EW values for stars in Coma Ber and that for stars in Praesepe  \cite[from][]{Nunez2024}. The two distributions are remarkably similar, with \halpha\ emission becoming evident around \bpminusrp\ = 2.3 mag, corresponding roughly to an M2 star, and increasing in similar ways for later spectral types in both clusters. 

Interestingly, the relative shift in color--period space seen for the mid and late M dwarfs in the two clusters seen in Figure~\ref{fig:CPDPraesepe} is not immediately apparent here. Our 2D KS tests confirmed that these EW distributions are not statistically distinct: when we considered the EWs for the 109 stars with \bpminusrp\ $>$ 1.84 (i.e., $\approx$M0 and later types) in Coma Ber and for the 268 in Praesepe, we found that p = 0.175. Repeating this test for with the 208 Hyads with \bpminusrp\ $>$ 1.84 and a measured EW compiled by \cite{Nunez2024}, we obtained p~=~0.091, once again indicating that the EWs for the two sets of stars are drawn from the same distribution. 

The D statistics underscore this conclusion. For the comparison of M dwarfs in Coma Ber and in Praesepe, D~=~0.142, with D$_{\rm critical}$~=~0.168. Meanwhile, D~=~0.167, with D$_{\rm critical}$~=~0.175, for the comparison of M dwarfs in Coma Ber and in the Hyades with \halpha\ EW measurements. In both cases, D~$<$~D$_{\rm critical}$, and we cannot reject the null hypothesis that the two sets of measurements are drawn from the same underlying distribution. 

This result suggests that the metallicity difference between Coma Ber and Praesepe and the Hyades makes little to no difference when examining the chromospheric activity of their stars, at least as traced by \halpha.

\subsection{\LLX\ distribution}\label{sec:Xrayresults}
Since the first detections of coronal X-rays by Einstein and ROSAT, studies of X-ray activity have used the ratio of the X-ray to the bolometric luminosity, \LLX, as this represents an estimate of the X-ray emission level that can be meaningfully compared across a range of masses \citep[e.g.,][]{Pallavicini1981,Randich1996}.

However, how \LLX\ should respond to variations in metallicity for stars of a given mass and age is still unclear. Complicating the picture is that observations of the solar corona have shown that certain elements (those with a relatively low first ionisation potential) are overabundant relative to what is measured in the solar photosphere, while elements with a higher first ionisation potential have the same abundance \citep{Feldman1992}. 

Observations of other stars have confirmed this difference in the abundances measured in photospheres and coronae, albeit with a dependence on the first ionization potential sometimes opposite to that seen in the Sun \citep[][see discussion in \citeauthor{Testa2010} \citeyear{Testa2010}]{Brinkman2001}.

What causes this observed relative depletion or enhancement of certain metals in the corona relative to the photosphere is not known. In their study of X-ray emission in Praesepe and the Hyades, \cite{Nunez2022} found a correlation between spectral type and coronal metal abundance: the K and M stars had very low
abundances, whereas F and G stars had higher abundances. This is consistent with other observations that suggest changes in the properties of the convective layer of the star can explain these variations in the measured coronal abundance \citep[e.g.,][]{Wood2012,Wood2013}, while others argue, based principally on high-resolution observations of the Sun, that magnetic activity itself causes these differences \citep[e.g.,][]{baker2020,lee2025}.

Because none of our Coma Ber spectra had a sufficiently large number of counts to allow for spectral analysis, we have no information on the composition of the coronae we detect in that cluster. With existing data we were limited to examining whether \LLX\ differs for clusters with different \emph{photospheric} metallicities.

\begin{figure}
\includegraphics{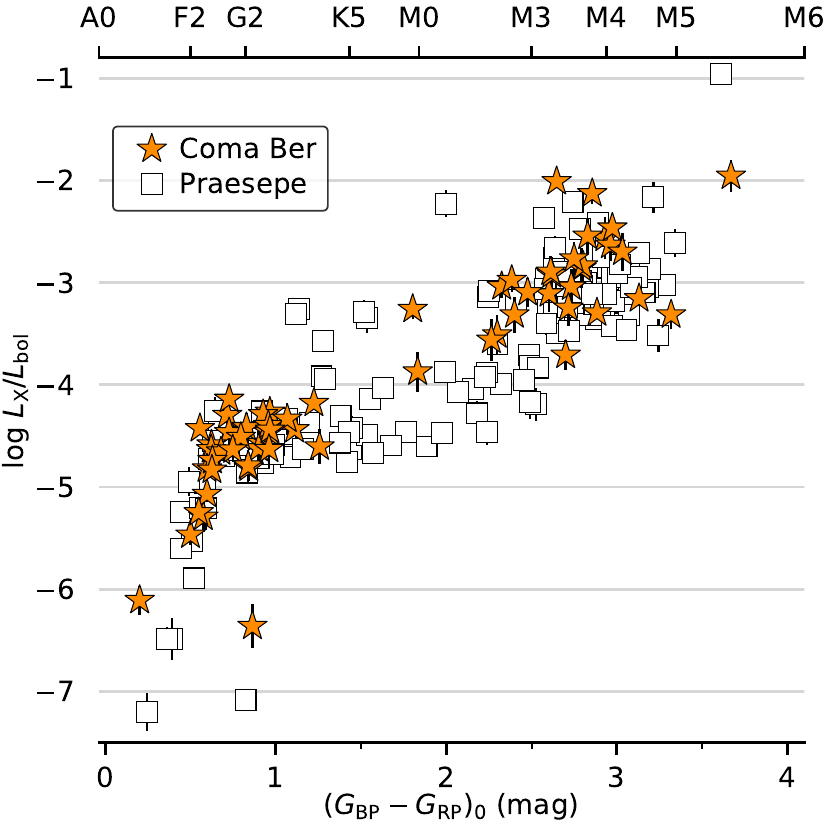}
\caption{\LLX\ as a function of \bpminusrp\ for likely single members of Coma Ber (orange stars) and of Praesepe \citep[white squares; data from][]{Nunez2024}. For most stars, the uncertainty in \LLX\ is smaller than the size of the symbol. While there are clear gaps in the X-ray coverage for Coma Ber, the measurements for the F and G stars on one hand and for the mid and late M dwarfs on the other appear consistent with those for these same stars in Praesepe. And, as with the \halpha\ measurements in Figure~\ref{fig:EW_color_Pr}, the distributions for those cooler stars do no appear offset from each other, unlike for their \Prot.}
\label{fig:X_color}
\end{figure}

In Figure~\ref{fig:X_color}, we plot the \LLX\ distribution for Coma Ber stars, once again including the data for Praesepe \citep[from][]{Nunez2024} for comparison. The hodgepodge nature of our  Coma Ber X-ray data is most obvious here, with two clumps of detections, one for F and G stars, and another for mid and late M dwarfs. The \LLX\ values for these stars appear largely to match those for their counterparts in Praesepe.

The sample of M dwarf X-ray emitters in Coma Ber is small, but for completeness we calculated KS statistics as above. A comparison of the 26 stars redder than \bpminusrp\ $=$ 1.84 with \LLX\ measurements in Coma Ber to the 85 in Praesepe returned p~=~0.109, with D~=~0.289 and a D$_{\rm critical}$~=~0.332. The same comparison with the 114 X-ray-emitting M dwarfs in the Hyades found p~=~0.066 and D~=~0.305, with D$_{\rm critical}$~=~0.322. In both cases, therefore, p~$>$~0.05 and D~$<$~D$_{\rm critical}$, and we cannot reject the null hypothesis that the measurements are drawn from the same underlying distribution. 

A final 2D KS test for the more massive (F through early K) rotators detected in both clusters returns a similar answer. For the 31 such stars in Coma Ber and 40 in Praesepe, p~=~0.268, and D~=~0.256 with D$_{\rm critical}$~=~0.354 (so D~$<$~D$_{\rm critical}$), and we cannot reject the null hypothesis that the measurements are drawn from the same underlying distribution.  

As with chromospheric activity, the metallicity difference between Coma Ber and its coeval clusters does not appear to impact the coronal activity of their stars as traced by X-rays. But a dedicated X-ray survey of Coma Ber---ideally one deep enough to allow for spectroscopy---is needed to strengthen this conclusion. 

\subsection{\LLH\ and \LLX\ dependence on \Ro}\label{sec:RossbyLL}

\begin{figure*}[!ht]
\includegraphics{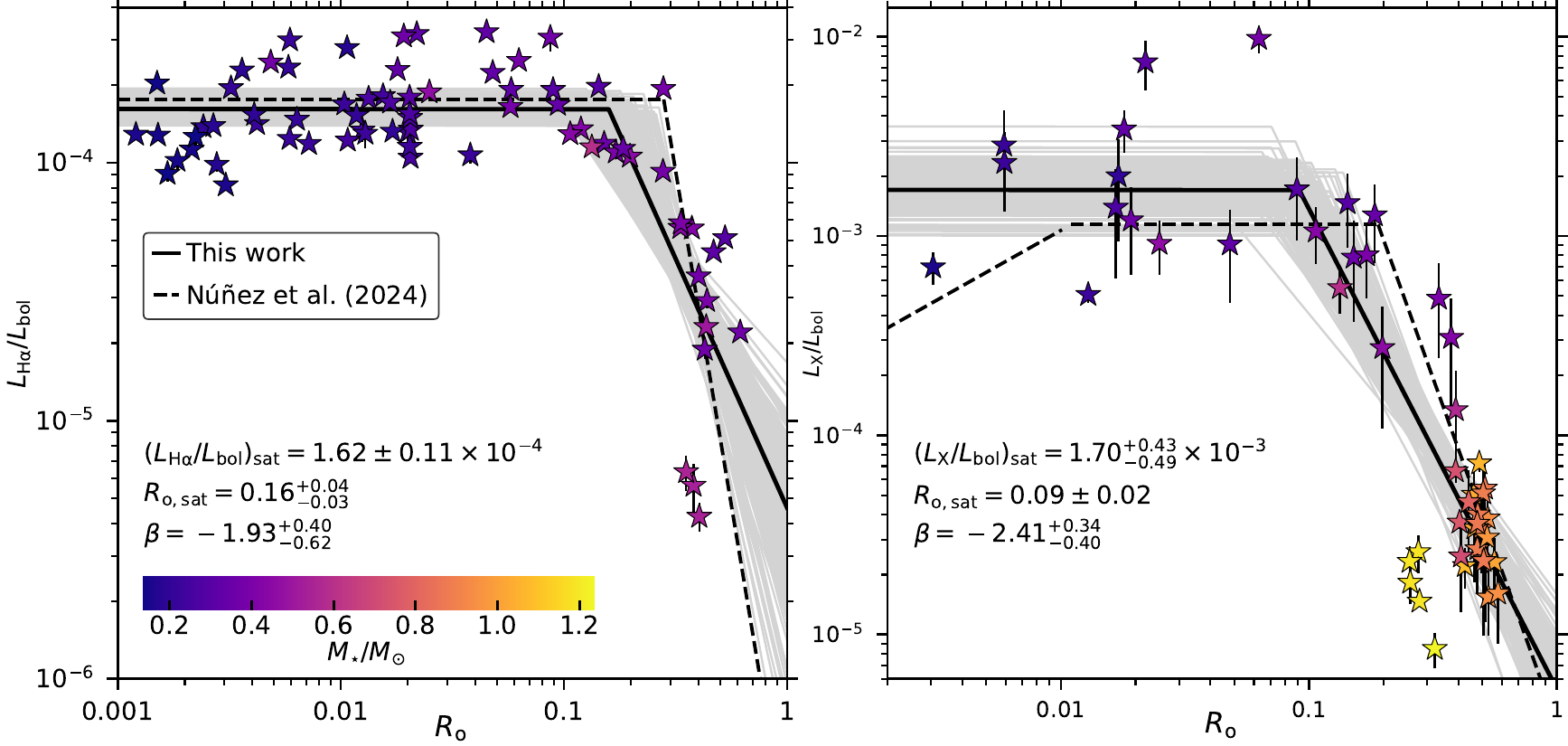}
\caption{\LLH\ (left) and \LLX\ (right) as a function of \Ro\ for likely single members of Coma Ber. Stars are color-coded by their $M_\star$ according to the colorbar in the left panel. The solid black line is the maximum \textit{a posteriori} MCMC fit, and the gray lines are 200 random samples from the posterior probability distributions. We assumed a flat saturated regime described by $(L_i/L_\mathrm{bol})_\mathrm{sat}$ (where $L_i$ is either $L_\mathrm{H\alpha}$ or \lx) and \Ro$_\mathrm{,sat}$, and an unsaturated regime described by a power-law of index $\beta$. The results of the fit for these three parameters are given in each panel. The dashed black lines are the results of a similar analysis for single members of Praesepe \citep{Nunez2024}, which in the right panel also includes a supersaturated regime (\Ro\ $\lapprox$~0.01). We show in \autoref{app:dists} the marginalized posterior probability distributions from the MCMC analysis for our fits.}
\label{fig:ro}
\end{figure*}

As is commonly done \citep[e.g.,][]{Randich2000b, Wright2011, Douglas2014, Nunez2015, Nunez2022, Nunez2024}, we parametrized the relation between \LLH\ and \Ro, and between \LLX\ and \Ro, as follows: for stars with \Ro\ $\leq$ \Ro$_\mathrm{,sat}$, activity is saturated (i.e., constant) and equal to $(L_i/L_\mathrm{bol})_\mathrm{sat}$, where $L_i$ is either $L_\mathrm{H\alpha}$ or \lx. Above \Ro$_\mathrm{,sat}$, activity declines as a power-law with index $\beta$, and is, therefore, unsaturated.

Functionally, this corresponds to
\begin{equation*}\label{eq:rossby}
  L_{i}/L_{\mathrm{bol}} = \left\{
  \begin{array}{l l} 
    (L_{i}/L_{\mathrm{bol}})_{\mathrm{sat}}
    & \quad \textrm{if $R_\mathrm{o}\le R_{\mathrm{o,sat}}$}\\
    C R_\mathrm{o}^{\beta} & \quad \textrm{if $R_\mathrm{o} > R_{\mathrm{o,sat}}$}
  \end{array} \right.
\end{equation*}
where $C$ is a constant. 

We used the Markov-chain Monte Carlo (MCMC) package \texttt{emcee} \citep{emcee} to fit our data with this three-parameter model, allowing for a nuisance parameter $f$ to account for underestimated errors.\footnote{See \url{https://emcee.readthedocs.io/en/develop/user/line/}. In including $f$ we followed the \texttt{emcee} implementation of \citet{Magaudda2020}.} We assumed flat priors for the three parameters, and used 300 walkers, each taking 5000 steps in their MCMC chain, to infer the maximum-likelihood parameters. The posterior distributions for each parameter and 2D correlations between pairs of parameters from each fit are included in \autoref{app:dists}. 

In Figure~\ref{fig:ro}, we show the \LLH\ and \LLX\ distributions as a function of \Ro, along with 200 random samples from the posterior distributions and the maximum \textit{a posteriori} model. Once again, we provide a comparison to the equivalent information for single members of Praesepe from \cite[][]{Nunez2024}. For this latter study, the assumed model also included a supersaturation regime, in which \LLX\ decays for \Ro\ $\lapprox$ 0.01.

The left panel of Figure~\ref{fig:ro}, for \LLH, shows clear agreement in the saturated regime between stars in Coma Ber and those in Praesepe. The characteristic \LLH\ = (1.61$\pm$0.11)$\times10^{-4}$ for saturated stars in Coma Ber, and (1.76$\pm$0.09)$\times10^{-4}$ in Praesepe. A comparison to the combined Praesepe and Hyades data presented in \cite{Nunez2024} underscores this similarity: for that sample of 312 likely single stars, the characteristic saturated \LLH = (1.65$\pm$0.06)$\times10^{-4}$. 

Interestingly, however, the transition from the saturated to the unsaturated regime does not occur at the same \Ro: for Coma Ber, this is at \Ro~=~0.16$^{+0.04}_{-0.03}$, while for Praesepe, it is at 0.28$^{+0.02}_{-0.03}$. (For the combined Praesepe and Hyades sample, this \Ro~=~0.29$\pm$0.01.)

An even more striking difference is apparent when considering the behavior of unsaturated stars: the power-law slope for Coma Ber's stars in this regime is $\beta = -1.93^{+0.40}_{-0.62}$, but $-5.19^{+1.32}_{-0.94}$ for Praesepe's. (Because $\beta$ is even steeper for unsaturated Hyads, the slope for the combined Praesepe and Hyades sample is even more discrepant; that $\beta = -5.85^{+0.81}_{-0.80}$.)

These differences in the \Ro$_\mathrm{,sat}$ and in the unsaturated $\beta$ are not surprising, however. As discussed in \cite{Nunez2024}, the quoted 1$\sigma$ uncertainties for \Ro$_\mathrm{,sat}$ are unrealistically small, as \Ro\ uncertainties are difficult to estimate and therefore not included when running the MCMC fit. This makes it unlikely that the \Ro$_\mathrm{,sat}$ obtained in different studies will agree statistically. 

Furthermore, the sample of unsaturated Praesepe and Hyades stars with measured \Prot\ and \halpha\ EWs is unusual: it is dominated by stars with $M_\star\gtrsim$~0.5~\Msun. As indicated by the color bar in the left panel of Figure~\ref{fig:ro}, the unsaturated Coma Ber stars tend to be less massive. Interestingly, the few unsaturated late K and early M massive stars in our sample (clumped at \Ro\ $\approx$ 0.4 and \LLH\ $\approx$ $5$$\times$$10^{-6}$) do imply a steeper $\beta$.

Echoing a point made by \cite{Nunez2024}, the difference between $\beta$ for unsaturated stars in Praesepe and for those in Coma Ber may be evidence for a steeper decline in chromospheric activity as a function of \Ro\ for partially convective stars compared to for fully convective stars. A more apt comparison for our results might therefore be those for the sample of field M dwarfs studied by \citet{Newton2017}, who found that \LLH~= (1.49$\pm$0.08)$\times10^{-4}$ in the saturated regime, \Ro$_\mathrm{,sat}$ = 0.21$\pm$0.02, and $\beta = -1.7$$\pm$$0.1$, values that are all compatible with those we derived for our Coma Ber stars. 

\begin{figure*}
\includegraphics{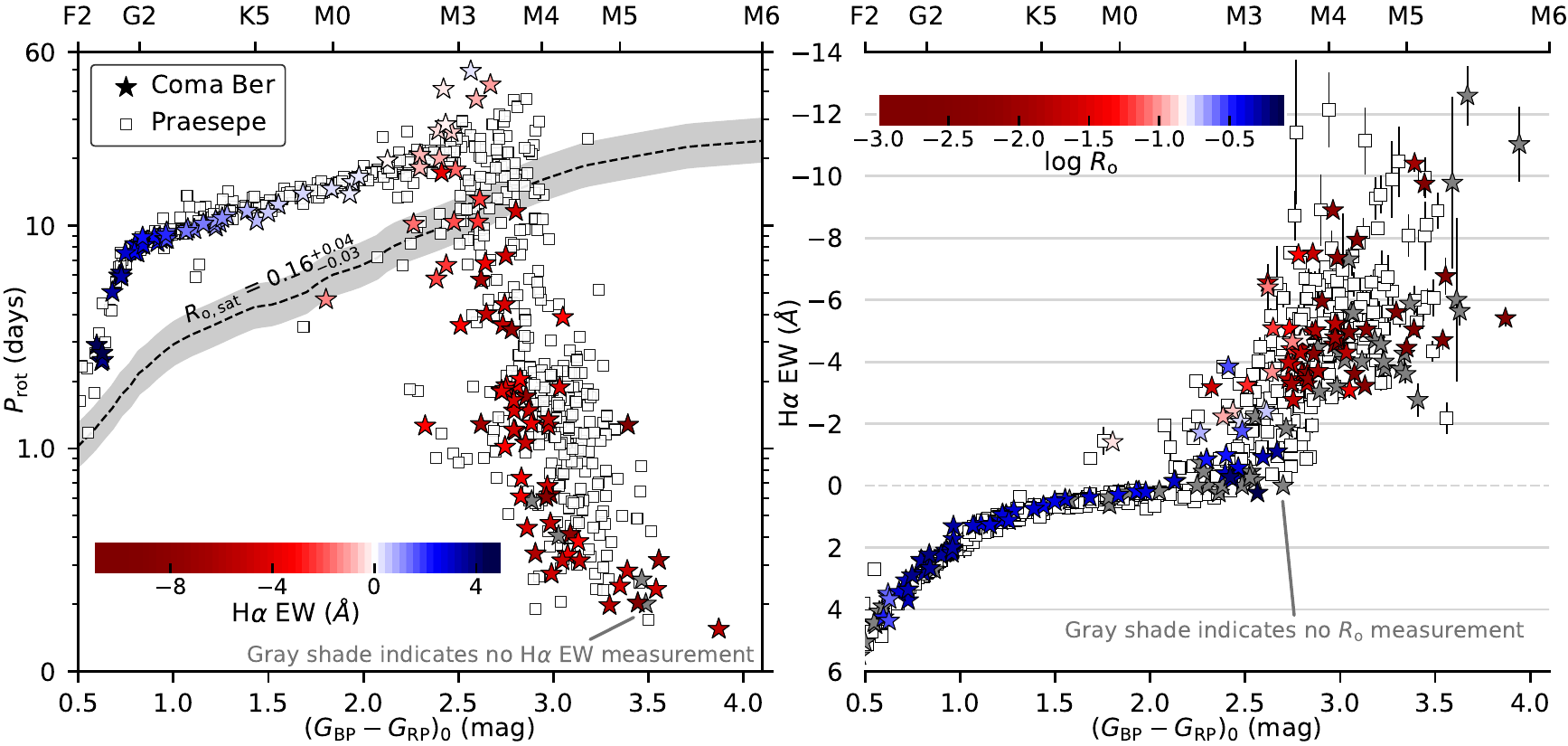}
\caption{Color--period (left panel) and color--EW (right panel) diagrams for likely single members of Coma Ber (colored stars) and Praesepe (white squares). In the left panel, Coma Ber stars are color-coded by their \halpha\ EW, and in the right panel, by their \Ro. Stars shaded in gray do not have an EW measurement (left panel) or a calculated \Ro\ (right panel). In the left panel, the dashed line and shaded area indicate \Ro$_\mathrm{,sat}$ and one standard deviation, respectively (see Figure~\ref{fig:ro}), which separates saturated stars (\Ro\ $<$ \Ro$_\mathrm{,sat}$) from unsaturated stars (\Ro\ $>$ \Ro$_\mathrm{,sat}$) in \LLH\ space. The white shade in the colorbar of the right panel corresponds to \Ro$_\mathrm{,sat}$ (log \Ro$_\mathrm{,sat} = -0.80\pm0.09$).}
\label{fig:massEW_Prot}
\end{figure*}

In the right panel of Figure~\ref{fig:ro}, by contrast, it is in the saturated regime that it is difficult to make meaningful comparisons to the data for Praesepe. There are few rapid rotators with measured \LLX\ in Coma Ber, presumably because those would be fainter sources, beyond the limits of what serendipitous observations or shallow all-sky surveys are likely to detect. The characteristic saturated \LLX\ = (1.71$^{+0.44}_{-0.47})$$\times10^{-3}$ is marginally consistent with that for single stars in Praesepe, for which \LLX\ = (1.14$\pm$0.12)$\times10^{-3}$, but this is based on only a dozen stars in Coma Ber. 

Once again, the values for \Ro$_\mathrm{,sat}$ are in disagreement, with \Ro$_\mathrm{,sat} = 0.09$$\pm$0.02 for Coma Ber and 0.19$\pm$0.02 for Praesepe. And the slope of the power-law for unsaturated stars is inconsistent with that for Praesepe members, with $\beta$ = $-2.43^{+0.34}_{-0.39}$ compared to $-3.48^{+0.34}_{-0.39}$. This last disagreement is somewhat less marked if considering the combined Praesepe and Hyades sample, for which $\beta = -3.18^{+0.20}_{-0.21}$. 

Collecting a larger sample of X-ray-detected rotators in Coma Ber will allow for more detailed analysis of the dependence of X-ray emission on rotation and comparisons between chromospheric and coronal behavior in low-mass stars.

\subsection{The key role of magnetic saturation}\label{sec:sat}

How can we reconcile our \Prot\ results, which suggest that metallicity does impact the relative distributions in color space of low-mass stars, with our activity results, which suggest that our measurements are largely insensitive to the metallicity difference between Coma Ber and Praesepe and the Hyades? Figure~\ref{fig:massEW_Prot} provides a possible explanation by highlighting the importance of magnetic saturation in this discussion. 

In the left panel, we plot the color--period diagram shown in the top left panel of Figure~\ref{fig:CPDPraesepe}, but now color-coding the Coma Ber stars by the strength of their \halpha\ line and indicating the transition between chromospherically unsaturated and saturated stars (the dashed line corresponds to \Ro~=~$0.16$). In the right panel, we replot Figure~\ref{fig:EW_color_Pr} and color-code the stars by their (log) \Ro. Both panels illustrate the expected relationship between rotation and magnetic activity: on the left, the fastest rotators have the strongest \halpha\ emission, and conversely, on the right, the stars with \halpha\ emission have the smallest \Ro.  

Beyond this, what Figure~\ref{fig:CPDPraesepe} reveals is that virtually all of the Coma Ber stars for which we measure significant \halpha\ emission are magnetically saturated: they are below the dashed line in the panel on the left, and they are plotted in red in the panel on the right. These stars' magnetic activity has plateaued, and is insensitive to changes in their \Prot\ and therefore \Ro. Given the agreement between the (\LLH)$_{\rm sat}$ levels shown in Figure~\ref{fig:ro}, it is therefore not surprising that a comparison between the chromospheric activity in stars in Coma Ber and in Praesepe and the Hyades does not reflect the differences seen in their \Prot\ distributions.

\section{Conclusion}
\label{sec:concl}

Coma Ber's proximity would suggest that it should be an obvious target for studies seeking to calibrate age-dependent stellar properties. But in the pre-Gaia era, identifying its members reliably was challenging, and the cluster's low-mass stars were not nearly as studied as their famous coeval cousins in Praesepe and the Hyades. 

By merging existing membership catalogs, some pre- and some post-Gaia DR2, and using Gaia DR3 to analyze the stars' properties, we constructed a Coma Ber catalog of 291 stars. We classified 262 of these stars as high-confidence members based on their parallaxes and either their UVW velocities or their proper motions and positions on the sky. The additional 29 stars are candidate members for which RV data are required to confirm their membership. Our catalog represents a significant increase in the number of Coma Ber stars relative to earlier catalogs, which typically included $<$200 stars.

With this updated catalog and spectroscopic measurements made with Keck HIRES, we determined Coma Ber's metallicity, finding that its metallicity is indistinguishable from solar. We also determined that the reddening in the direction of Coma Ber is almost exactly 0~mag, unsurprising given the cluster's proximity. 

We then used the appropriate PARSEC isochrones to fit Coma Ber's CMD, and estimated that it is 675$\pm$100~Myr, making its stars coeval with those in the well-studied benchmarks Praesepe and the Hyades. However, Coma Ber's metallicity is 0.2 dex lower than that of those two clusters, making comparisons between stellar properties in the three particularly valuable for exploring e.g., how metallicity impacts the rotation--activity relation.

By combining literature \Prot\ with our own measurements based on TESS and ZTF photometry, we assembled a catalog of 161 \Prot\ for Coma Bar's low-mass stars. Of these, 137 (85\%) are new measurements, enabling a far more complete description of the rotational behavior of stars in the cluster that extends well into the M dwarfs. A comparison of this \Prot\ distribution and that for Praesepe shows that the slowly rotating sequences of late F, G, and K stars for the two clusters are remarkably similar, despite the 0.2~dex difference in metallicity between them. It appears that differences on this scale are insufficient to significantly impact the rotational evolution of these stars, or that whatever differences there may have been, they have been erased by 700 Myr. For stars that are within a few tenths of a dex of solar metallicity, applying gyrochronological relations to  estimate ages based on their colors and \Prot\ can be done even with imperfect knowledge of their metallicities. 

By contrast, the \Prot\ distributions appear to differ for stars with \bpminusrp\ $\gapprox$~2, corresponding to spectral types later than $\approx$M0. We conducted 2D KS tests comparing the \Prot\ distribution for these stars in Coma Ber to those in Praespe and the Hyades, finding that they are statistically distinct. The 0.2 dex difference in metallicity is sufficient to impact the apparent timescale for spin down for M dwarfs. An M3 star in Praesepe is more likely than its cousin in Coma Ber to have spun down sufficiently to have joined the slow-rotating sequence defined by the G and K stars in the cluster. In this color regime, applying a standard gyrochronological relation without knowledge of the stars' metallicity is likely to provide a misleading age, as the overall \Prot\ distribution for the metal-rich stars makes the population appear older than it is.

Stars of the same mass but different metallicities will have different colors, so that this comparison of the \Prot\ distributions in color space, while appropriate for observers, cannot on its own address how the rotational evolution of stars of the same mass (rather than of the same color) is impacted by changes in the metallicity. 

When we consider the Coma Ber and Praesepe \Prot\ distributions in mass space, the difference with the results in color space is striking. The \Prot\ for the lowest-mass cluster members are now well-matched, an agreement confirmed by a 2D KS test that indicates that the \Prot\ are drawn from the same underlying distribution. 

A plausible explanation for the disappearance of the difference in the \Prot\ distributions for the later-type stars as we move from color to mass space is that, as discussed below, in all three clusters most of these stars are magnetically saturated. This implies that the strength of their magnetic field no longer depends on their rotation rate, and by extension, the torque exercised by the stellar winds on these stars is insensitive to the difference in their metallicities. In mass space, saturation erases the impact metallicity may have on spin down.

Meanwhile, if one focuses on the \Prot\ for the $>$0.6 \Msun\ (late F, G, and K) stars, there is a small offset between the two slow-rotating sequences that is most obvious for the K stars. These stars in Coma Ber have not spun down as much as their more metal-rich counterparts in Praesepe, consistent with both theoretical predictions and observations of stars in the Kepler field.

We characterized chromospheric activity in Coma Ber by measuring the EW of the \halpha\ line for nearly 250 members. Our comparison in color space to the EW distribution for stars in Praesepe found that the two are remarkably similar, with \halpha\ emission becoming evident around \bpminusrp\ = 2.3 mag, corresponding to an M2 star, and increasing in similar ways for later spectral types in the two clusters. 

The offset in color--period space seen for the mid and late M dwarfs in the two clusters is not apparent here, and our 2D KS tests confirmed that these EW distributions are not statistically distinct. This result suggests that in color space, the metallicity difference between Coma Ber and Praesepe (and the Hyades) makes little to no difference when examining the chromospheric activity of their stars as traced by \halpha.

We also collected X-ray detections for $\approx$100 Coma Ber members from serendipitous source catalogs and all-sky surveys. A comparison of the \LLX\ distributions for the M dwarfs in Coma Ber and in Praesepe suggests that the two are drawn from the same underlying distribution. As with our measurements of chromosphetic activity, the metallicity difference between Coma Ber and its coeval clusters does not appear to impact the coronal activity of their stars as traced by X-rays. 

Combining our activity measurements and \Prot\ data, we examined the dependence on \Ro\ of \LLH\ and \LLX. Magnetically saturated stars in \LLH\ in Coma Ber appear to behave very similarly to those in Praesepe and the Hyades, with \LLH\ $\approx$ 1.6$\times$10$^{-4}$ for all stars with an \Ro\ smaller than a threshold \Ro\ that is lower for Coma Ber (\Ro = $0.16^{+0.04}_{-0.03}$ versus 0.28$^{+0.02}_{-0.03}$ for Praesepe and 0.29$\pm0.01$ for the combined Praesepe and Hyades sample). Although potentially interesting, this difference in \Ro$_\mathrm{,sat}$ is hard to interpret, as uncertainties in \Ro\ are difficult to estimate and the quoted values are almost certainly underestimated.

As for unsaturated stars,  where \LLH\ increases as \Ro\ decreases (i.e., the stars spin faster), the power-law dependence on \Ro\ appears to be different in Coma Ber. The slope defined by Coma Ber's stars in this regime is much shallower than for their cousins in Praesepe and the Hyades ($\beta = -1.90^{+0.40}_{-0.62}$ for Coma Ber, but $-5.19^{+1.32}_{-0.94}$ for Praesepe and $-5.85^{+0.81}_{-0.80}$ for the combined Praesepe and Hyades sample). 

However, the unsaturated stars in the three clusters are not the same: in Praesepe and the Hyades, these are mostly $\gtrsim$0.5~\Msun\ stars, while in Coma Ber, these stars tend to be less massive. The difference in the slopes may therefore be evidence for a steeper decline in chromospheric activity as a function of \Ro\ for partially convective stars than for fully convective stars. 

We found that our Coma Ber results for \Ro$_\mathrm{,sat}$ and $\beta$ were compatible with values found for field M dwarfs, which may be a more appropriate sample for this comparison.

The X-ray detections in Coma Ber are primarily of slowly rotating stars. While we were able to estimate values for $\beta$, for \Ro$_\mathrm{,sat}$, and for \LLX$_{\rm , sat}$, interpreting these results is difficult given the poor X-ray coverage we have for the cluster. A dedicated X-ray survey of Coma Ber, one that is deep enough to allow for spectroscopy and that targets many more rotators, is needed before the cluster can contribute meaningfully to discussions about the dependence of X-ray emission on rotation or comparisons between chromospheric and coronal behavior in low-mass stars.

Magnetic saturation is the key to reconciling our \Prot\ results, which indicate that metallicity does impact the observed distributions in color space of low-mass stars, with our activity results, which suggest that our measurements are largely insensitive to the metallicity difference between Coma Ber and Praesepe and the Hyades. Virtually all of the Coma Ber stars for which we measure significant \halpha\ emission are magnetically saturated, as are the \halpha\ emitters in Praesepe and the Hyades. These stars' magnetic activity has plateaued, and is insensitive to changes in their \Prot---and therefore \Ro---that are expected to occur for stars of a given mass but differing metallicities. It is therefore not surprising that a comparison between the chromospheric activity in stars in Coma Ber and in Praesepe and the Hyades does not reflect the differences seen in their \Prot\ distributions.

\section*{Acknowledgments}
We thank the anonymous reviewer for comments that improved the manuscript.

We thank Elliana S.~Abrahams, Justin Anderson, Victoria DiTomasso, David Fierroz, Sofia Lawsky, Justin Rupert, John Thorstensen, Minzhi (Luna) Wang, and Jos\'e Manuel Zorrilla Matilla for their assistance in obtained spectra for this work at the MDM Observatory.  

We are grateful for two Sigma Xi Grants-in-Aid of Research to S.~Douglas (G201510151605844) and A.~N\'u\~nez (G2016100191859320) that supported our MDM spectroscopic campaign.

We thank Sean Matt and members of his group for helpful discussions.

M.A.A.~acknowledges support from a Fulbright U.S.~Scholar grant co-funded by the Nouvelle-Aquitaine Regional Council and the Franco-American Fulbright Commission and from a Chrétien International Research Grant from the American Astronomical Society. M.A.A.~also acknowledges support provided by NASA through TESS grants 80NSSC23K0369 (program ID G05157), 80NSSC23K0370 (program ID G05028), and 80NSSC24K0502 (program ID G06164). 
J.L.C.~acknowledges support provided by the NSF through grant AST-2009840 and by NASA through TESS grant 80NSSC24K0501 (program ID G06163). 
A.N.~acknowledges support provided by the NSF through grant 2138089. B.S.~acknowledges support provided by NASA TESS grant 80NSSC22K0299 (program ID G04217). 
M.K.~acknowledges support provided by the NSF under grant AST-2205736, and NASA under grants 80NSSC22K0479, 80NSSC24K0380, and 80NSSC24K0436.

This work has made use of data from the European Space Agency (ESA) mission Gaia,\footnote{\url{https://www.cosmos.esa.int/gaia}} processed by the Gaia Data Processing and Analysis Consortium (DPAC).\footnote{\url{https://www.cosmos.esa.int/web/gaia/dpac/consortium}} Funding for the DPAC has been provided by national institutions, in particular the institutions participating in the Gaia Multilateral Agreement. This research also made use of public auxiliary data provided by ESA/Gaia/DPAC/CU5 and prepared by Carine Babusiaux.

This work is based on observations obtained at the MDM Observatory, operated by Dartmouth College, Columbia University, Ohio State University, Ohio University, and the University of Michigan. The authors are honored to be permitted to conduct astronomical research on Iolkam Du'ag (Kitt Peak), a mountain with particular significance to the Tohono O'odham.

This work has benefited from the public data released from the Guoshoujing Telescope (Large Sky Area Multi-Object Fiber Spectroscopic Telescope, or LAMOST), a National Major Scientific Project built by the Chinese Academy of Sciences. Funding for the project has been provided by the National Development and Reform Commission. LAMOST is operated and managed by the National Astronomical Observatories, Chinese Academy of Sciences.

Funding for the Sloan Digital Sky Survey IV has been provided by the Alfred P. Sloan Foundation, the U.S. Department of Energy Office of Science, and the Participating Institutions. SDSS acknowledges support and resources from the Center for High-Performance Computing at the University of Utah.

The Pan-STARRS1 Surveys (PS1) and the PS1 public science archive have been made possible through contributions by the Institute for Astronomy, the University of Hawaii, the Pan-STARRS Project Office, the Max-Planck Society and its participating institutes, the Max Planck Institute for Astronomy, Heidelberg and the Max Planck Institute for Extraterrestrial Physics, Garching, The Johns Hopkins University, Durham University, the University of Edinburgh, the Queen's University Belfast, the Harvard-Smithsonian Center for Astrophysics, the Las Cumbres Observatory Global Telescope Network Incorporated, the National Central University of Taiwan, the Space Telescope Science Institute, the National Aeronautics and Space Administration under Grant No. NNX08AR22G issued through the Planetary Science Division of the NASA Science Mission Directorate, the National Science Foundation Grant No. AST-1238877, the University of Maryland, Eotvos Lorand University (ELTE), the Los Alamos National Laboratory, and the Gordon and Betty Moore Foundation.

PyRAF is a product of the Space Telescope Science Institute, operated by AURA for NASA. The Image Reduction and Analysis Facility, IRAF, is distributed by the NOAO, which are operated by the AURA, Inc., under cooperative agreement with the NSF.

This paper includes data collected by the TESS mission, which are publicly available from the Mikulski Archive for Space Telescopes (MAST). Funding for the TESS mission is provided by NASA’s Science Mission directorate. 

Based on observations obtained with the Samuel Oschin Telescope 48-inch and the 60-inch Telescope at the Palomar Observatory as part of the Zwicky Transient Facility project. ZTF is supported by the National Science Foundation under Grants No. AST-1440341 and AST-2034437 and a collaboration including current partners Caltech, IPAC, the Weizmann Institute for Science, the Oskar Klein Center at Stockholm University, the University of Maryland, Deutsches Elektronen-Synchrotron and Humboldt University, the TANGO Consortium of Taiwan, the University of Wisconsin at Milwaukee, Trinity College Dublin, Lawrence Livermore National Laboratories, IN2P3, University of Warwick, Ruhr University Bochum, Northwestern University and former partners the University of Washington, Los Alamos National Laboratories, and Lawrence Berkeley National Laboratories. Operations are conducted by COO, IPAC, and UW. 

This research has made use of data obtained from the 4XMM XMM-Newton Serendipitous Source Catalog compiled by the 10 institutes of the XMM-Newton Survey Science Centre selected by ESA, and of data obtained from the Chandra Source Catalog, provided by the Chandra X-ray Center (CXC) as part of the Chandra Data Archive.

This work is based on data from eROSITA, the soft X-ray instrument aboard SRG, a joint Russian-German science mission supported by the Russian Space Agency (Roskosmos), in the interests of the Russian Academy of Sciences represented by its Space Research Institute (IKI), and the Deutsches Zentrum für Luft- und Raumfahrt (DLR). The SRG spacecraft was built by Lavochkin Association (NPOL) and its subcontractors, and is operated by NPOL with support from the Max Planck Institute for Extraterrestrial Physics (MPE). The development and construction of the eROSITA X-ray instrument was led by MPE, with contributions from the Dr. Karl Remeis Observatory Bamberg \& ECAP (FAU Erlangen-Nuernberg), the University of Hamburg Observatory, the Leibniz Institute for Astrophysics Potsdam (AIP), and the Institute for Astronomy and Astrophysics of the University of Tübingen, with the support of DLR and the Max Planck Society. The Argelander Institute for Astronomy of the University of Bonn and the Ludwig Maximilians Universität Munich also participated in the science preparation for eROSITA.

\vspace{5mm}
\facilities{eROSITA, CXO, Gaia, Hiltner (OSMOS, Modspec), LAMOST, Pan-STARRS, PO:1.2m (ZTF), Sloan, TESS, XMM}

\software{astropy \citep{astropyIII}, astroquery \citep{astroquery}, emcee \citep{emcee}, IDL Astronomy User's Library \citep{IDLastro}, Matplotlib \citep{matplotlib}, NumPy \citep{numpy}, PHEW \citep{PHEW}, PypeIt \citep{Prochaska2020,Prochaska2020zndo}, PyRAF \citep{pyraf}, PySpecKit \citep{Ginsburg2011}, SciPy \citep{scipy}, TESScut, unpopular \citep{Hattori2022}, wdwarfdate \citep{wdwarfdate}}

\bibliography{main.bib}{}
\bibliographystyle{aasjournal}

\appendix 
\section{Analysing the WD Gaia DR3 4001560041148002432 }\label{app:wds}

\begin{figure}[th!]
\centerline{\includegraphics[trim=.75cm .5cm 1.75cm .5cm, clip=True, width=.5\columnwidth]{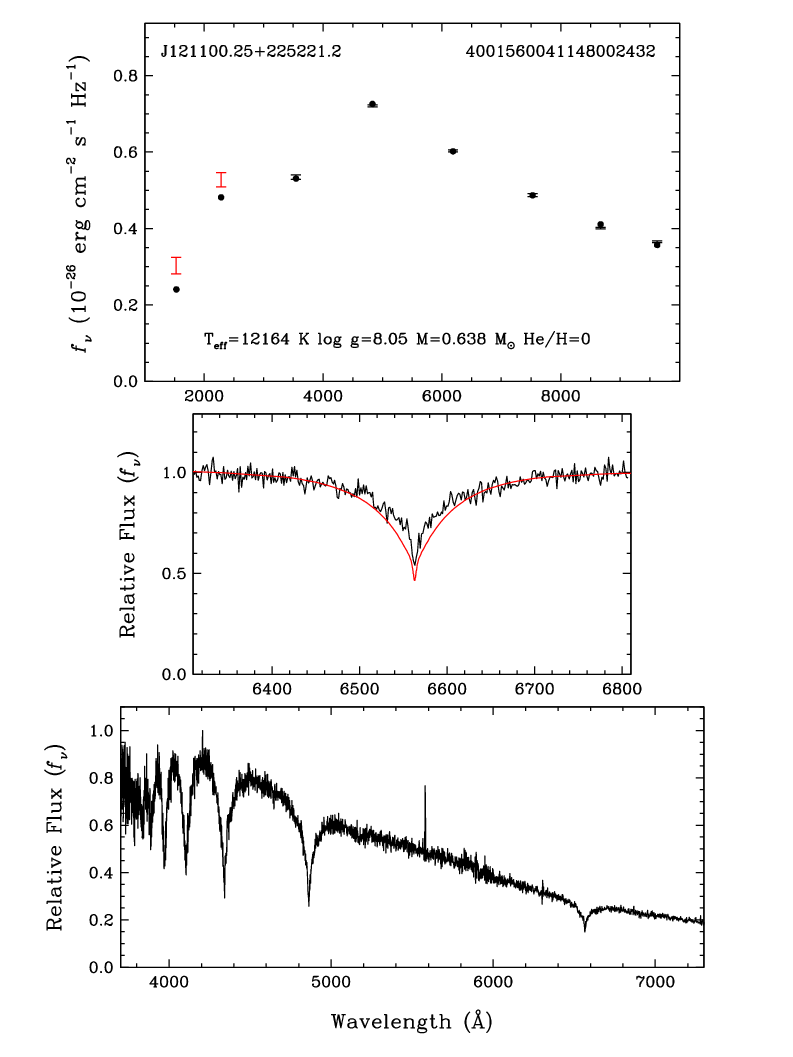}}
\caption{\textit{Top---}Spectral energy distribution for \object{Gaia DR3 4001560041148002432}. The error bars are the photometry available for the WD, including GALEX far- and near-ultraviolet (in red, and not used for fitting), SDSS u, and Pan-STARSS griz. The black dots are the derived flux values from the best-fit model to the ugriz data, and the numbers shown correspond to the  parameters for this best-fit pure-hydrogen model atmosphere. These parameters were used to create a model spectrum for the WD. \textit{Middle---}The model spectrum is plotted in red, with the LAMOST spectrum in black, for a window centered on the \halpha\ line. The match is clearly poor, with the model predicting a far broader absorption feature than what is observed. \textit{Bottom---}The LAMOST spectrum for the WD.}
\label{fig:app_wd1}
\end{figure}

We followed \citet{kilic2020} in analyzing the available photometry for \object{Gaia DR3 4001560041148002432}. The observed magnitudes were converted into average fluxes using the appropriate zero points, and were compared to average synthetic fluxes calculated from a pure-hydrogen atmosphere model. The top panel in Figure~\ref{fig:app_wd1} shows the average observed fluxes (error bars) along with the best fit synthetic fluxes (black dots) for this WD. The resulting stellar parameters were then used to generate a model spectrum, shown in red in the middle panel for the region around the \halpha\ line. The photometric fit clearly predicts a much broader line than that obtained by LAMOST. The bottom panel shows the full LAMOST spectrum, for reference. 

\begin{figure}[!h]
\begin{center}
\centerline{\includegraphics[trim=8.75cm 1.15cm 0.5cm 18.cm, angle=270, clip=True, width=.3\columnwidth]{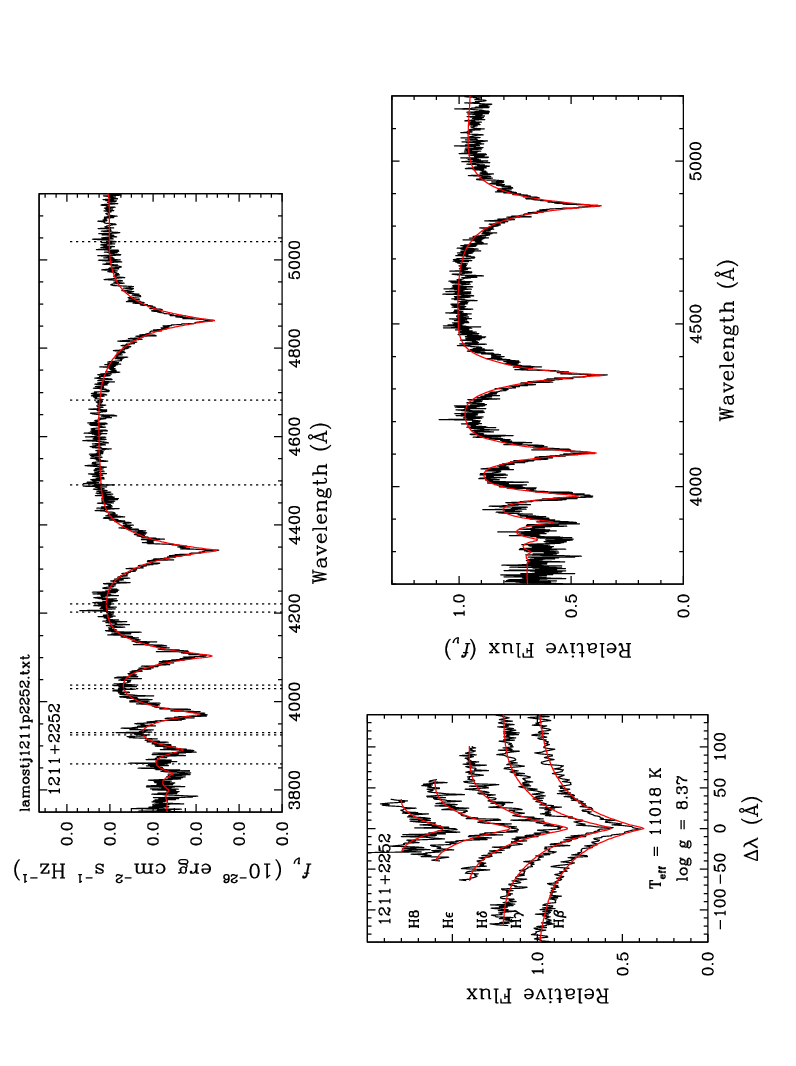}}
\caption{Cutouts of the WD's Balmer lines fit using a 1D model for the WD atmosphere. The three-dimensional corrections in both \teff\ and log~g given in \citet{tremblay13} have been applied to take into account hydrodynamical effects. This fit produces significantly different values for the WD's fundamental properties. } 
\label{fig:app_wd2}
\end{center}
\end{figure}

We therefore fit the normalized Balmer lines in the LAMOST spectrum using a one-dimensional (1D) model for the WD atmosphere \citep[][see Figure~\ref{fig:app_wd2}]{bergeron1992}. The stellar properties that gave the best fit were then corrected to three dimensions. The resulting spectroscopically derived values differ significantly from those obtained from the photometry. The WD's mass is 0.638 \Msun\ and its $\tau_{\rm cool}$~=~408~Myr photometrically, and 0.838 \Msun\ and its $\tau_{\rm cool}$~=~814~Myr spectroscopically, reflecting an almost 1150 K difference in the \teff\ derived for the star. We give the values for the WD's parameters as determined from these two approaches in Table~\ref{tbl:differences}. Further observations are required to determine the nature of this discrepancy, which e.g., may be due to the WD being an unresolved double-degenerate system.

\section{Marginalized Posterior Probability Distributions for the MCMC Analysis of Our \Ro--\LLH\ and \Ro--\LLX\ Models}\label{app:dists}

\begin{figure*}[!hb]
\centerline{\includegraphics[width=.85\columnwidth]{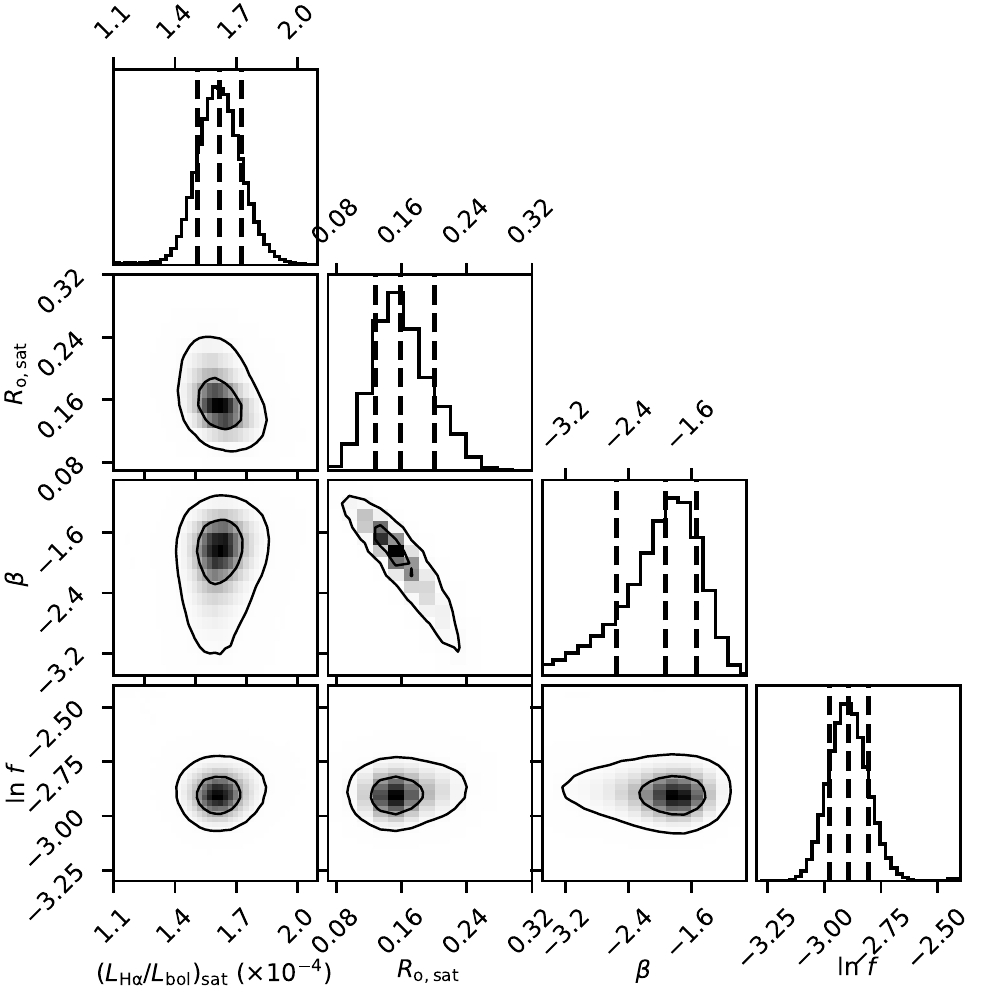}}
\caption{Marginalized posterior probability distributions from the MCMC analysis of our \Ro--\LLH\ model using \texttt{emcee} for single members in Coma Ber. The parameter values of the \textit{a posteriori} model are the peaks of the 1D distributions; the vertical dashed lines approximate the median and 16$^{\rm th}$, 50$^{\rm th}$, and 84$^{\rm th}$ percentiles. The 2D distributions illustrate covariances between parameters; the contour lines approximate the 1$\sigma$ and 2$\sigma$ levels of the distributions.}
\label{fig:posteriorsHa}
\end{figure*}

\begin{figure*}
\centerline{\includegraphics[width=.85\columnwidth]{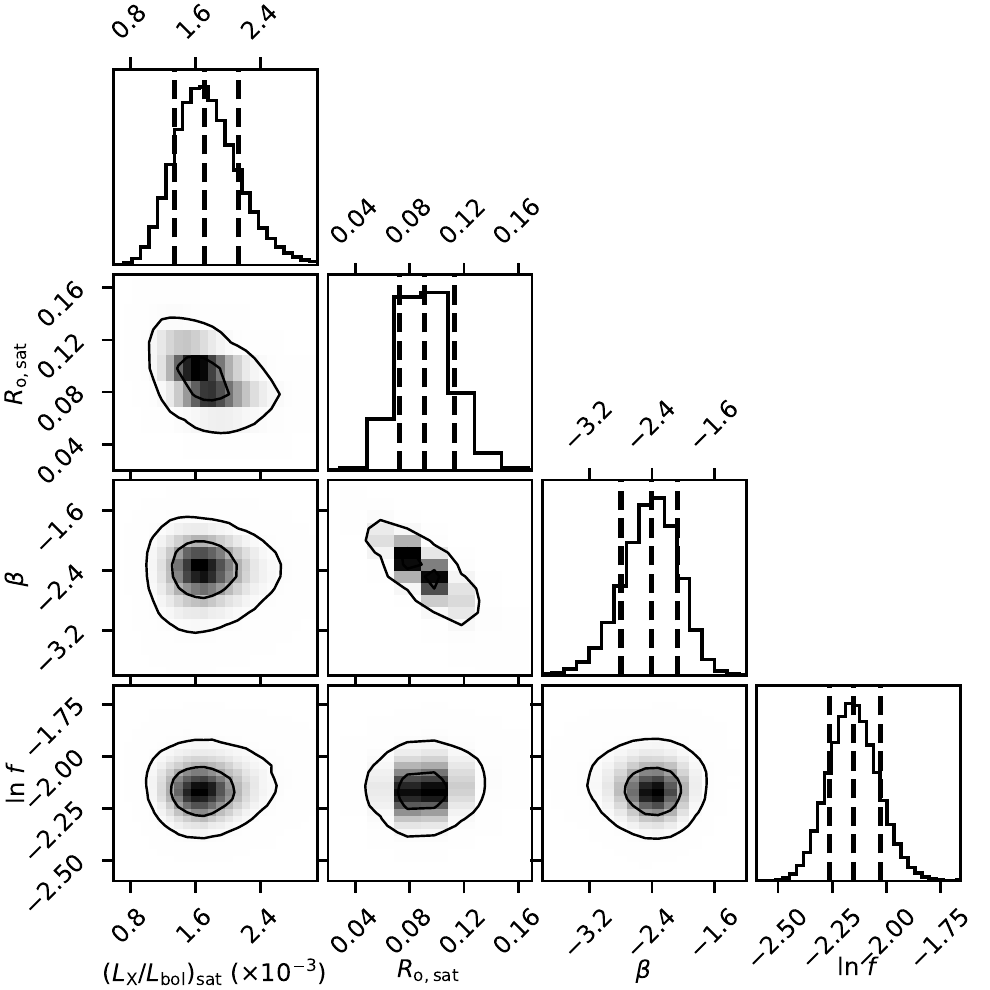}}
\caption{Same as Figure \ref{fig:posteriorsHa}, but for our \Ro--\LLX\ model.}
\label{fig:posteriorsX}
\end{figure*}

We present the marginalized posterior probability distributions from the MCMC analysis we performed on our sample of single members of Coma Ber (see Section \ref{sec:RossbyLL}). Figure \ref{fig:posteriorsHa} shows the marginalized posterior probability distribution for the \Ro--\LLH\ model, and Figure \ref{fig:posteriorsX}, for the \Ro--\LLX\ model.

\begin{deluxetable}{lcccc}
\centering 
\tablecaption{Properties for \object{Gaia DR3 4001560041148002432}, as derived from our two different approaches \label{tbl:differences}}
\tablehead{
\colhead{Fit to...}  & 
\colhead{\teff} &
\colhead{Mass} &
\colhead{log g} &
\colhead{$\tau_{cool}$}\\[-0.1in]
\colhead{} &
\colhead{(K)} & 
\colhead{(\Msun)} &
\colhead{(dex)} &
\colhead{(Myr)}
}

\startdata
photometry & 12,164$\pm$108 & 0.638$\pm$0.09 & 8.054$\pm$0.011 & 408.2$\pm$11.5\\
spectroscopy & 11,018$\pm$188 & 0.838$\pm$0.043 & 8.372$\pm$0.066 & 813.7\\
\enddata
\end{deluxetable}

\end{document}